\documentclass[universe,article,accept,pdftex,oneauthor]{Definitions/mdpi} 

\firstpage{1} 
\pubvolume{10}
\issuenum{7}
\articlenumber{291}
\pubyear{2024}
\copyrightyear{2024}
\externaleditor{Academic Editors: Matt Visser, Jessica Santiago, Sebastian Schuster and Ana Alonso-Serrano}
\datereceived{9 May 2024} 
\daterevised{23 June 2024} 
\dateaccepted{4 July 2024} 
\datepublished{6 July 2024} 
\hreflink{https://doi.org/10.3390/\linebreak universe10070291} 

\usepackage{bm}
\usepackage{amssymb}
\usepackage{amsfonts}
\usepackage{mathrsfs}
\usepackage{amsmath}
\usepackage{amsthm}
\usepackage{xcolor}
\usepackage{bbold}
\usepackage{braket}
\usepackage{physics}
\usepackage{graphicx}
\usepackage{makecell}
\DeclareMathAlphabet{\mathbbold}{U}{bbold}{m}{n}

\newcommand{\be}{\begin{equation}}
	\newcommand{\ee}{\end{equation}}
\newcommand{\bea}{\begin{eqnarray}}
	\newcommand{\eea}{\end{eqnarray}}
\def\bml{\begin{subequations}}
	\def\blea{\bml\begin{eqnarray}}
		\def\eml{\end{subequations}}
	\def\elea{\end{eqnarray}\eml}

\newcommand{\mc}[1]{\mathcal{#1}}

\newcommand{\Tren}{T^{\text{ren}}}
\newcommand{\Tsplit}{\mathbb T^{\text{split}}}
\newcommand{\Tfin}{T^{\text{fin}}}
\newcommand{\coin}[1]{\left[\!\!\left[#1\right]\!\!\right]}
\newcommand{\nord}[1]{{:}#1{:}}
\def\Rmax{R_{\text{max}}}

\Title{Wormhole Restrictions from Quantum Energy Inequalities}

\TitleCitation{Wormhole Restrictions from Quantum Energy Inequalities}

\Author{Eleni-Alexandra Kontou} 

\AuthorNames{Eleni-Alexandra Kontou}
\AuthorCitation{Kontou, E.-A}

\address[1]{%
Department of Mathematics, King’s College London, Strand, London WC2R 2LS, UK; eleni.kontou@kcl.ac.uk}

\abstract{Wormhole solutions, bridges that connect different parts of spacetime, were proposed early in the history of General Relativity. Soon after, it was shown that all wormholes violate classical energy conditions, which are non-negativity constraints on contractions of the stress--energy tensor. Since these conditions are violated by quantum fields, it was believed that wormholes can be constructed in the context of semiclassical gravity. But negative energies in quantum field theory are not without restriction: quantum energy inequalities (QEIs) control renormalized negative energies averaged over a geodesic. Thus, QEIs provide restrictions on the construction of wormholes. This work is a review of the relevant literature, thus focusing on results where QEIs restrict traversable wormholes. Both `short' and `long' (without causality violations) wormhole solutions in the context of semiclassical gravity are examined. A new result is presented on constraints on the Maldacena, Milekhin, and Popov `long' wormhole from the recently derived doubled smeared null energy condition. }

\keyword{energy conditions; wormholes; semiclassical gravity; quantum energy inequalities; averaged null energy condition} 

\begin{document}

\section{Introduction}

Wormholes, bridges that connect different parts of spacetime, are dominant in science fiction. The~idea of ``shortcuts'' that can reduce the time of travel between stars and galaxies is an appealing one. But~what does physics say about them? Interestingly, the~general theory of relativity, the~most complete theory of gravity so far, allows their existence. Indeed, one can imagine any spacetime they desire and, solving the Einstein equation in reverse, find out the matter needed to construct~it. 

Along with the development of general relativity came the idea of energy conditions: restrictions on the type of matter that can exist to avoid the unphysical ones. Initially, energy conditions were used exactly as assumptions, thus expressing desirable qualities such as the positivity of energy density. Perhaps their most famous use is the singularity theorems of Penrose~\cite{Penrose:1964wq} and Hawking~\cite{Hawking:1966sx}. 

{Coincidentally in the~same year that Penrose submitted his Nobel prize winning manuscript, Epstein, Glaser, and Jaffe, showed that the non-negativity of energy is incompatible with quantum field theory~\cite{Epstein:1965zza}. But~quantum field theory does not allow unlimited amounts of negative energy either. Ford~\cite{Ford:1978qya} was the first to suggest a bound on the timelike averaged renormalized energy and flux of quantum fields. This was the first \textit{quantum energy   inequality} (QEI).}

Unlike the original energy conditions, QEIs are derived directly from a quantum field theory, thus exhibiting a property of known matter and not functioning as assumptions. Their strength is also their weakness; the derived QEIs are not general bounds obeyed by all fields. Remarkably though, the~energy density of all free fields on flat and curved spacetimes exhibits the same behavior when averaged over a timelike geodesic: less negative energy is allowed when averaged over longer times. Less is known for self-interacting fields, as there are minimal results and only in two-dimensions (see Section~\ref{sec:qeis} for the relevant references).

What about null geodesics? Unlike the timelike ones, when averaging over a finite segment of a null geodesic, there exist states with unlimited negative energy~\cite{Fewster:2002ne}. Only averaging over an entire null geodesic restores the non-negativity of the renormalized null energy, thus giving one of the most important energy conditions: the averaged null energy condition (ANEC). 

The ANEC over achronal null geodesics that satisfies the (semiclassical) Einstein equation is free of counterexamples in semiclassical gravity. More importantly, it is sufficient to rule out the existence of causality violations in asymptotically flat spacetimes~\cite{Graham:2007va}. 

While the achronal ANEC seems to prohibit wormholes as ``shortcuts'' \endnote{A `short' wormhole can be modified to act as a time machine, as  discussed in Section~\ref{sub:time}.}, a~different kind of wormhole still seems possible: the~`long' wormhole. This is a wormhole where it takes longer for an observer to travel through the throat than the outside spacetime. Such a wormhole might seem useless as an interstellar travel medium or a time machine, but~it is still physically interesting as a different spacetime topology. `Long' wormholes circumvent the problem of the achronal ANEC, as there are no complete achronal null geodesics passing through~them. 

In this case, null QEIs could provide restrictions. To~go around the counterexample of Ref.~\cite{Fewster:2002ne}, two different approaches have been suggested: (1) imposing a momentum or length cutoff, which leads to what is called the smeared null energy condition (SNEC) \cite{Freivogel:2018gxj}, and (2) averaging over two null directions, thus leading to the double smeared null energy condition (DSNEC) \cite{Fliss:2021phs}.

In this work, I review the different types of energy conditions and QEIs and discuss the constraints they provide to wormhole geometries. I particularly discuss the constraints to a long wormhole, the~one proposed by Maldacena, Milekhin, and Popov~\cite{Maldacena:2018gjk} by the SNEC and the DSNEC. The~constraint provided by the DSNEC is a new~result. 

The review is organized as follows: In Section~\ref{sec:poinwise}, I briefly review the original or pointwise energy conditions. In~Section~\ref{sec:qeis}, I discuss QEIs (timelike and null) and~discuss the ANEC and its derivations. In~Section~\ref{sec:wormholes}, I introduce wormhole geometries and their relationship with energy condition violations. In~Sections~\ref{sec:short} and \ref{sec:long}, I analyze the restrictions from QEIs and the achronal ANEC on short and long wormholes, respectively. I conclude in Section~\ref{sec:discussion} with a summary and~discussion.

\subsection*{Conventions}
Unless otherwise specified, we work in $n$ spacetime dimensions and use the metric signature $(-,+,\ldots,+)$.  We work mainly in units where $\hbar=c=1$, except~some parts where $G_N$ is set to $1$ or where dimensions are restored. The~D'Alembertian operator with respect to the metric $g$ is defined as $\Box_g := -g^{\mu \nu} \nabla_\mu \nabla_\nu$. The~Riemann curvature tensor of $g$ is 
\vspace{-6pt}{}
\be
R(X,Y)Z = \nabla_X \nabla_Y Z - \nabla_Y \nabla_X Z - \nabla_{[X,Y]}Z \,,
\ee
and the Ricci tensor $R_{\mu \nu}$ is its $(1,3)$-contraction. The~Einstein Equation is 
\vspace{-6pt}{}
\be
\label{eqn:Einstein}
G_{\mu \nu}=8\pi G_N T_{\mu \nu} \,.
\ee
The convention used for the metric, the~Riemann tensor, and the Einstein equation is the $[+,+,+]$ according to Misner, Thorne, and Wheeler~\cite{MTW}.

\section{Pointwise Energy~Conditions}
\label{sec:poinwise}
\unskip

\subsection{Brief~Overview}

Energy conditions, as~statements on the non-negativity of energy density and pressure of the universe, appeared early on in the history of general relativity. Historically, they initially constrained components of a perfect fluid stress--energy tensor. Later, they appeared as statements on contractions of a general stress--energy tensor on any spacetime time. These conditions are called \text{pointwise}. 

They are famously used in the singularity theorems of  Penrose~\cite{Penrose:1964wq} and Hawking~\cite{Hawking:1966sx}. There, what is used is what we call the \textit{geometric form} of the energy conditions: a restriction on contractions of the Ricci tensor. This form is connected with the \textit{physical form}, which includes the stress--energy tensor with the Einstein equation. Even though the two forms are often mixed in the literature, they are distinct. One is about the matter content, and the other is about the spacetime geometry. Only if we have a self-consistent solution of the Einstein Equation, which is difficult even classically, can we discuss a full equivalence of the two~forms. 

The pointwise conditions were introduced as assumptions or suggestions to what a ``reasonable'' matter model should look like. 
 As~they are not derived, counterexamples exist even for classical fields, as we will see in the next subsection. This makes them subject to fair criticism~\cite{Curiel:2014zba, Barcelo:2002bv}; however, they are still used in a variety of~situations. 

Here, we will discuss three pointwise energy conditions: the weak energy condition (WEC), the~strong energy condition (SEC), and the null energy condition (NEC). The~latter is the hardest to violate and thus probably the most~important. 

The WEC requires that, for every future-pointing timelike vector, $t^\mu$
\vspace{-6pt}{}
\be
\label{eqn:WEC}
T_{\mu \nu} t^\mu t^\nu \geq 0 \,.
\ee
For a perfect fluid stress--energy tensor, this requirement becomes
\vspace{-6pt}{}
\be
\rho \geq 0 \qquad \text{and} \qquad \rho+P \geq 0 \,,
\ee
so the energy density $\rho$ is non-negative at any spacetime point. Additionally, the pressure cannot be so negative that it dominates the energy density. Both requirements seem intuitive for classical fields. The~geometric form of the WEC,
\vspace{-6pt}{}
\be
G_{\mu \nu} t^\mu t^\nu \geq 0 \,,
\ee
is not easy to interpret as a requirement for the spacetime curvature, as the Einstein tensor does not have a direct geometric~meaning. 

The SEC requires
\be
\label{eqn:EED}
\left(T_{\mu \nu} - \frac{T}{n-2}g_{\mu \nu}\right) t^\mu t^\nu \geq 0\,,
\ee
where we assume $n>2$. The~quantity on the left was named~\cite{Brown:2018hym} \textit{effective energy density} (EED), which was inspired by Pirani~\cite{Pirani:1956tn}, who calls this quantity the effective density of gravitational mass. Despite the name, the~SEC does not imply the WEC, although~it is generally more easily violated. For~a perfect fluid stress--energy tensor, it becomes
\vspace{-6pt}{}
\be
\rho+P \geq 0 \qquad \text{and} \qquad  (n-3)\rho +(n-1) P \geq 0 \,.
\ee
The  geometric form of the SEC is called the \textit{timelike convergence condition},
\be
\label{eqn:timelikeconvergencecondition}
R_{\mu \nu} t^\nu t^\nu \geq 0 \,,
\ee
and it is the form more commonly used. It implies that a irrotational timelike geodesic congruence locally converges. It was used in the Hawking singularity theorem~\cite{Hawking:1966sx}, which predicts singularities in cosmological~scenarios.

The physical form of the NEC is
\vspace{-6pt}{}
\be
T_{\mu \nu} \ell^\mu \ell^\nu \geq 0 \,,
\ee
where $\ell^\mu$ is a null vector. 
For a perfect fluid stress--energy tensor, it becomes
\vspace{-6pt}{}
\be
\rho+P \geq 0 \,,
\ee
so it is easy to see that the WEC implies the NEC. Its geometric form is called the \textit{null convergence condition},
\be
\label{eqn:nullconvergencecondition}
R_{\mu \nu} \ell^\mu \ell^\nu \geq 0 \,,
\ee
and it implies that an irrotational null geodesic congruence locally converges. It has been used in several classical relativity theorems, including the Penrose singularity theorem~\cite{Penrose:1964wq} and the Hawking area theorem~\cite{Hawking:1971vc}. It also plays an important role in constructing and sustaining wormhole geometries, as we will~see.

These pointwise energy conditions are summarized in Table~\ref{tab:pointwise}. 

\begin{table}[H]
\caption{A summary of the three pointwise conditions discussed: WEC, SEC, and~NEC.}
		\label{tab:pointwise}
		\begin{tabularx}{\textwidth}{m{1.5cm}<{\centering}CCC} \toprule
			\textbf{Condition} &  \textbf{Physical Form} & \textbf{Geometric Form} & \textbf{Perfect Fluid} \\ \midrule
			\makecell{WEC} & $T_{\mu \nu} t^\mu t^\nu \geq 0$ &	$G_{\mu \nu} t^\mu t^\nu \geq 0$ & \makecell{$\rho \geq 0$ and\\ $\rho+P \geq 0$} \\ \midrule
			\makecell{SEC} & $(T_{\mu \nu} -\frac{T}{n-2}g_{\mu \nu})t^\mu t^\nu \geq 0$ &	$R_{\mu \nu} t^\mu t^\nu \geq 0$ & \makecell{$\rho + P \geq 0$ and\\ $(n-3)\rho+(n-1)P \geq 0$} \\ \midrule
			\makecell{NEC} & $T_{\mu \nu} \ell^\mu \ell^\nu \geq 0$ &	$R_{\mu \nu} \ell^\mu \ell^\nu \geq 0$ & $\rho + P \geq 0$ \\ 
			\bottomrule
		\end{tabularx}
\end{table}
\unskip

\subsection{Example: The Klein--Gordon~Field}

Here, I will briefly review the minimally and nonminimally coupled classical Klein--Gordon field and discuss the three pointwise energy conditions introduced in the previous~subsection. 

We consider the massive minimally coupled classical scalar field $\phi$ with the field equation
\vspace{-6pt}{}
\be
\label{eqn:KleinGordon}
(\Box_g+m^2)\phi=0 \,,
\ee
where $m$ has dimensions of inverse length. The~Lagrangian is
\vspace{-6pt}{}
\be
L[\phi]=\frac{1}{2} \left( (\nabla \phi)^2+m^2 \phi^2 \right) \,,
\ee
where $(\nabla \phi)^2=g^{\mu \nu} \nabla_\mu \phi \nabla_\nu \phi$. Varying the action with respect to the metric gives the classical stress--energy tensor
\be
\label{eqn:stresstensorcl}
T_{\mu \nu}=\nabla_\mu \phi \nabla_\nu \phi-\frac{1}{2} g_{\mu \nu} ((\nabla \phi)^2+ m^2) \,.
\ee

It is easy to see that the NEC is obeyed in this case, as well as the WEC, because $t^\mu t_\mu$ is negative for a timelike vector. For~the SEC, we have 
\vspace{-6pt}{}
\be
T_{\mu \nu} -\frac{T}{n-2}g_{\mu \nu} = \nabla_\mu \phi\nabla_\nu \phi +\frac{1}{n-2}g_{\mu \nu}m^2 \phi^2 \,.
\ee
Then, we can see that the SEC holds only in the massless case, but~it can be violated \mbox{when $m>0$.}

Perhaps a more interesting case is that of the nonminimally coupled scalar field. The~field equation in this case is
\vspace{-6pt}{}
\be
\label{eqn:eom}
\left( \Box_g+m^2+\xi R \right) \phi=0 \,,
\ee
while the Lagrangian is 
\be
\label{eqn:graction}
L[\phi]=\frac{1}{2}((\nabla \phi)^2+\xi R\phi^2+m^2 \phi^2) \,,
\ee
where $\xi$ is the dimensionless coupling constant to the curvature. The~conformal coupling in $n$ dimensions is 
\be
\xi_c=\frac{n-2}{4(n-1)} \,.
\ee
Thus, for four-dimensions, $\xi_c=1/6$. The~stress--energy tensor is
\vspace{-6pt}{}
\be
\label{eqn:stressten}
T_{\mu \nu}=(\nabla_\mu \phi)(\nabla_\nu \phi)-\frac{1}{2} g_{\mu \nu}(m^2 \phi^2+(\nabla \phi)^2)+\xi(-g_{\mu \nu} \Box_g -\nabla_\mu \nabla_\nu+G_{\mu \nu})\phi^2 \,.
\ee
We should note that the stress--energy tensor is different from the minimal coupling one, even for zero curvature. Using the identity
\vspace{-6pt}{}
\be
\phi \Box_g \phi=\frac{1}{2}\Box_g \phi^2 -(\nabla \phi)^2 \,,
\ee
we can write
\vspace{-6pt}{}
\bea
\label{eqn:tmn}
T_{\mu \nu}&=&(1-2\xi)(\nabla_\mu \phi)(\nabla_\nu \phi)-\frac{1}{2}(1-4\xi) g_{\mu \nu}(m^2 \phi^2+\xi R \phi^2+(\nabla\phi)^2) \nonumber \\
&&-2\xi\left(\phi \nabla_\mu \nabla_\nu \phi+\frac{1}{2} R_{\mu \nu}\phi^2\right) \,.
\eea
The stress--tensor contracted with two null vectors is
\vspace{-6pt}{}
\be
\label{eqn:nullenergy}
T_{\mu \nu} \ell^\mu \ell^\nu=(1-2\xi)(\ell^\mu \nabla_\mu \phi)(\ell^\nu \nabla_\nu \phi)-2\xi\left(\phi (\ell^\mu \ell^\nu \nabla_\mu \nabla_\nu \phi) +\frac{1}{2} R_{\mu \nu} \ell^\mu \ell^\nu \phi^2 \right) \,.
\ee
So, we see that the NEC can be violated, and thus, the WEC and the SEC can be violated~too. 

However, the~Equation~\eqref{eqn:nullenergy} is not exactly the physical form of the NEC, as the curvature appears in the stress tensor. We can define an {\it effective stress tensor} \cite{Barcelo:2000zf} by separating the curvature terms from the field terms in the Einstein equation. For~the purposes of constraining classical spacetimes allowed by the Einstein equation, this form of the stress tensor is more useful, as it is directly connected to the geometry. So, using the Einstein Equation
and Equation~\eqref{eqn:stressten}, we have
\vspace{-6pt}{}
\be
\label{eqn:eeeff}
G_{\mu \nu}=8\pi G_N T^{\text{eff}}_{\mu \nu} \,,
\ee
where
\bea
\label{eqn:Teff}
T^{\text{eff}}_{\mu \nu}&=&\frac{1}{1-8\pi G_N \xi \phi^2}\bigg((\nabla_\mu \phi)(\nabla_\nu \phi)-\frac{1}{2} g_{\mu \nu}\left(m^2 \phi^2+(\nabla \phi)^2\right) \nonumber\\
&&\qquad \qquad \qquad+\xi(-g_{\mu \nu} \Box_g -\nabla_\mu \nabla_\nu)\phi^2 \bigg) \,.
\eea
Then,
\be
\label{eqn:nullenergyeff}
T^{\text{eff}}_{\mu \nu} \ell^\mu \ell^\nu=\frac{1}{1-8\pi G_N \xi \phi^2}\left((\ell^\mu \nabla_\mu \phi)(\ell^\nu \nabla_\nu \phi)-\xi(\ell^\mu \ell^\nu \nabla_\mu \nabla_\nu)\phi^2 \right) \,.
\ee
From this form, we understand that the effective NEC can be violated in this case for both positive and negative $\xi$. We also see that for $\xi>0$, there is a critical value of the field $\phi$, where the effective stress tensor changes sign. A~change of sign of the coefficient of the stress tensor means the change of the sign of the Einstein equation. These values are considered unphysical~\cite{Barcelo:2000zf, Fliss:2023rzi}.

\section{Quantum Energy~Inequalities}
\label{sec:qeis}

In the previous section, we saw that there exist classical fields which violate the pointwise energy conditions. The~situation is even worse in quantum field theory (QFT). In~1964, Epstein, Glaser, and Jaffe~\cite{Epstein:1965zza} showed that the positivity of the energy density is generally incompatible with quantum field theory (QFT). They proved that any (nonzero) observable on Minkowski spacetime with a vanishing vacuum expectation value has a nonzero probability for both positive and negative measurement~values. 

However, quantum fields cannot admit unlimited negative energies. Ford~\cite{Ford:1978qya} was the first to constrain the time average of the quantized expectation value of energy density and flux. His motivation was to avoid the violation of the generalized second law of black hole thermodynamics. If~the energy flux $F$ is constrained by an inequality of the form
\vspace{-6pt}{}
\be
|F| \gtrapprox \frac{1}{\tau^2} \,,
\ee
where $\tau$ is the time that it lasts, then the generalized second law is preserved. This is an example of a quantum energy inequality (QEI). Generally, they are bounds on the renormalized expectation value of a contraction of the stress--energy tensor averaged over a spacetime region. Unlike their first derivation, QEIs are mainly derived directly from QFT, thus choosing a specific type of quantum field on a flat or curved~spacetime. 

In this section, I introduce QEIs with an emphasis on those averaged over null geodesics. I begin with a a brief introduction to algebraic quantization that is used to derive most~QEIs.

\subsection{Algebraic~Quantization}

To state the relevant QEIs, we need to quantize and renormalize the classical stress--energy tensor. We will only refer to free minimally coupled scalars with the stress--energy tensor of Equation~\eqref{eqn:stresstensorcl}. However, QEIs have been derived for other types of fields, which are results that will only be mentioned briefly~here. 

We begin with a brief step-by-step description of the method for the reader not familiar with the algebraic quantization and Hadamard renormalization of the~stress tensor:
\begin{itemize}
	\item 
	In the algebraic approach, the fields are smeared with a smooth function and are elements of an algebra obeying properties that include the field equation and canonical commutation relations.
	\item 
	The states $\psi$ are linear maps from the algebra to the field of complex numbers, while the two-point function $W_\psi$ is a bilinear map between the space of two functions and the field of complex numbers. 
	\item 
	In quantum field theory, products of fields---for~example, the two-point function---are divergent when one tries to naively compute their expectation values. But~an important aspect of the process is the choice of states: Hadamard states have a known singularity structure. Thus, it is sufficient to subtract these singularities from the two-point function to the correct order. The~Hadamard parametrix is a bidistribution $H_{(k)}(x,x')$ that includes all the singularities up to order $k$.  
	\item 
	The renormalization proceeds using point spliting: the classical stress tensor is written as an operator acting on two different points $(x,x')$.  After~we have subtracted the singularities, we act on  the two-point function with the point split operator  $\mathbb{T}_{\mu\nu'}^{\text{split}}(x,x')$, and we have a smooth object at coincident limit:
\vspace{-6pt}{}
\be
 \lim_{x \to x'} \left[ \mathbb{T}_{\mu\nu'}^{\text{split}}(W_\psi-H_{(k)})(x,x') \right] \,.
 \ee
 \item 
 This, up~to renormalization ambiguities, corresponds to the $\langle \Tren_{\mu \nu}\rangle_\psi$---the~object used instead of the classical stress tensor $T_{\mu \nu}$ in QEIs. 
\end{itemize}

Now, we will add a few more details and the relevant references to the previous discussion. As~mentioned, we follow the algebraic approach (see~\cite{KhavkineMorettiaqft} for a review), which will be reviewed here briefly. We start by introducing a unital $*$-algebra $\mathscr{A}(\mathcal{M})$
 on the manifold $\mathcal{M}$ so that self-adjoint elements of $\mathscr{A}(\mathcal{M})$ are observables of the theory. The~algebra is generated by elements $\Phi(f)$, where $f \in C_0^\infty(\mathcal{M})$, which is a~function that is differentiable for all degrees of differentiation. The~elements $\Phi(f)$  represent smeared quantum fields and obey the following~relations:
\begin{itemize}
	\item
	\textbf{Linearity} \\
	The map $f\rightarrow\Phi(f)$ is complex linear;
	\item
	\textbf{Hermiticity} \\
	$\Phi(f)^* = \Phi(\overline{f})$ \qquad $\forall f \in C^{\infty}_0(\mathcal{M})$;
	\item
	\textbf{Field Equation} \\
	$\Phi((\Box_g+m^2) f) = 0$  \qquad $\forall f \in C^{\infty}_0(\mathcal{M})$;
	\item
	\textbf{Canonical Commutation Relations} \\
	$\left[\Phi(f), \Phi(h)\right] = iE (f,h)\mathbbold{1}$  \qquad $\forall f,h \in C^{\infty}_0(\mathcal{M})$.
\end{itemize} 
The functions $f,h$ belong to the space of smooth functions of compact support from $\mathcal{M}$ to $\mathbb{C}$, and 
\be
E(x,y)=E^{A}(x,y)-E^R(x,y) 
\ee
is an antisymmetric bidistribution equal to the difference of the advanced $E^{A}(x,y)$ and retarded $E^R(x,y)$ Green functions of $\Box_g+m^2$, which exist due to global hyperbolicity of the~spacetime.  

The states of the theory are described by linear maps $\psi:\mathscr{A}(\mathcal{M})\to\mathbb{C}$ that are normalized and positive. For~a given state, $\psi$, the~two-point function is a bilinear map $W_{\psi}$ between $ C^{\infty}_0(\mathcal{M})\times C^{\infty}_0(\mathcal{M})$, and $\mathbb{C}$ is defined by $W_{\psi}(f,h)=\psi(\phi(f)\phi(h))$. Not all the states have physically desirable properties. Additionally, general curved spacetimes do not have a preferred vacuum state. The~class of states we consider are the Hadamard states~\cite{KhavkineMorettiaqft, moretti2021global}, which have a known two-point function singularity structure close to that of the Minkowski vacuum. For~technical definitions of Hadamard states and the wavefront set definition due to Radzikowski, see Ref.~\cite{Radzikowski:1996pa}.

We proceed to renormalize the stress--energy tensor following the prescription of Hollands and Wald
~\cite{Hollands:2001nf, Hollands:2004yh}. The~local time-ordered products of fields that compose the quantized stress tensor should obey a number of axioms that express their desired properties. Those properties include locality, continuity, analyticity, symmetry of the factors, and~unitarity. 

First, let us define the point split stress--energy operator
\vspace{-6pt}{}
\be
\Tsplit_{\mu \nu'}(x,x')=\nabla^{(x)}_\mu \otimes \nabla^{(x')}_{\nu'}-\frac{1}{2} g_{\mu \nu'}(x,x')g^{\lambda \rho'}(x,x') \nabla^{(x)}_\lambda \otimes \nabla^{(x')}_{\rho'}+\frac{1}{2} m^2 g_{\mu \nu'}(x,x') \mathbbold{1} \otimes \mathbbold{1} \,,
\ee
where $g_{\mu \nu'}(x,x')$ is the parallel propagator implementing the parallel transport of vectors along the unique geodesic connecting $x$ and $x'$.

Now, let us consider a Hadamard state $\psi$ with a two-point function $W_{\psi}$. To~construct the expectation value of the renormalised stress--energy tensor in that state $\langle T_{\mu\nu}^{\text{ren}}\rangle_{\psi}(x)$, we first introduce the Hadamard parametrix: a bidistribution that encodes the singularity structure of the two-point function of Hadamard states, which is expressed as an infinite series~\cite{Decanini:2005eg}. In~$n=4$ dimensions, the Hadamard parametrix takes the form
\vspace{-6pt}{}
\be
W_{\psi}(x,x')=\frac{U(x,x')}{4 \pi^2\sigma(x,x')}+V(x,x')\log (\frac{\sigma(x,x')}{2\ell^2})+\text{smooth}\,,
\ee
where $U$ and $V$ are real and symmetric smooth functions constructed from the metric and its derivatives, $\sigma$ is the squared geodesic distance or Synge's world function, and~$\ell$ is a particular length~scale.

Now, we subtract the Hadamard parametrix $H_{(k)}$, where the subscript $(k)$ means terms up to order $k$, to~remove the singularities of the two-point function and take the \mbox{coincidence limit:}
\be
\langle T^{\text{fin}}_{\mu\nu}\rangle_{\psi}(x)=\coin{ g_{\nu}^{\,\,\,\,\nu'}(x,x')\mathbb{T}_{\mu\nu'}^{\text{split}}(W_{\psi}-H_{(k)})(x,x') }\,.
\ee
We denote the coincidence limit of a generic bidistribution, $\mathcal{B}(x,x')$, as
\vspace{-6pt}{}
\be
[\![\mathcal{B}(x,x')]\!]=\lim\limits_{x\rightarrow x'} \mathcal{B}(x,x') \,.
\ee
The renormalized stress tensor on curved spacetimes will always have finite ambiguities: 
\vspace{-6pt}{}
\be
\label{eqn:localcurv}
\langle \Tren_{\mu \nu} \rangle_\psi (x)=\langle \Tfin_{\mu \nu}  \rangle_\psi-Q(x) g_{\mu \nu}(x)+C_{\mu \nu}(x) \,.
\ee
Those take the form of a state-independent conserved local curvature term $C_{\mu \nu}$ that vanishes in Minkowski space. Here, $Q$ is a term introduced by Wald~\cite{Wald:1978pj} to preserve the conservation of the stress--energy tensor.
We also define the difference of $\langle \Tren_{\mu \nu} \rangle$ between two Hadamard states $\psi$ and $\psi_0$, which is smooth at the coincidence limit:
\vspace{-6pt}{}
\be
\label{eqn:diff}
\langle :T_{\mu\nu}: \rangle_{\psi}:=\coin{ g_{\nu}^{\,\,\,\,\nu'} \mathbb{T}_{\mu\nu'}^{\text{split}}(W_{\psi}-W_{\psi_0})}=\langle T_{\mu\nu}^{\text{ren}} \rangle_{\psi}-\langle T_{\mu\nu}^{\text{ren}} \rangle_{\psi_0}\,.
\ee

\subsection{Overview of Quantum Energy~Inequalities}
\label{sub:QEIs}

Now, I will introduce QEIs, denoting $\rho$, as~some contraction on the stress--energy tensor, thus renormalized according to the procedure described~above. 

The most general form of a QEI takes the form
\vspace{-6pt}{}
\be
\langle \rho(f) \rangle_\psi \geq -\langle \mathcal{Q}(f) \rangle_\psi \,,
\ee
where $f$ is a real test function, and the operator $\mathcal{Q}$ is allowed to be unbounded. $\psi$ is the class of states for which the inequality holds; here, we only consider Hadamard states. When the operator $\mathcal{Q}(f)$ is a multiple of the identity, so $\langle \mathcal{Q}(f)\rangle_\psi=Q(f)$, we call the QEI~state independent.

Another important categorization of QEIs are the \textit{absolute} and \textit{difference} ones. We call difference QEIs the ones where we bound the smooth difference between the expectation values of $\rho$ of two Hadamard states, as defined in Equation~\eqref{eqn:diff}:
\vspace{-6pt}{}
\be
\langle \nord{\rho} (f)\rangle_\psi \geq -\langle \mathcal{Q}_{\psi_0}(f) \rangle_\psi \,.
\ee
Then, the bound can depend on both the reference state $\psi_0$ and the state of interest $\psi$. However, if it only depends on the reference state, it is a state-independent bound. If~instead, we renormalize using the Hadamard parametrix, the~QEI is called absolute. If~the reference state is the massless Minkowski vacuum, the two kinds~coincide.

Finally, we can make the distinction between averaging over a spacetime path (geodesic or not) and averaging over a spacetime volume. We will call the first type a \textit{worldline} QEI and the second a \textit{worldvolume} QEI. In~the second category, we also have averaging over domains with less than the spacetime dimension (e.g., area bounds). 

Perhaps one the most famous QEIs is for the energy density of a free scalar field on Minkowski spacetime. The~inequality takes the form
\vspace{-6pt}{}
\be
\label{eqn:QEIminflat}
\int_\gamma dt f^2(t) \langle \nord{T_{\mu \nu}t^\mu t^\nu} \rangle_{\psi} \geq -\frac{1}{16\pi^2} \int dt (f''(t))^2 \,,
\ee
where the average is over a timelike geodesic $\gamma$. The~stress--energy tensor is normal-ordered, and $\psi$ is the class of Hadamard states. It was first derived by Ford and Roman for Lorentzian functions~\cite{Ford:1994bj} and later generalized by Fewster and Eveson~\cite{Fewster:1998pu} for general compactly supported functions. We should note here that, while functions of compact support are considered in general to be derivations of QEIs to keep the lower bounds finite, in~applications, often functions that decrease very fast such as Gaussians are~used. 

By rescaling the sampling function, we have
\vspace{-6pt}{}
\be
\label{eqn:fscale}
f_{\tau}(t)=\frac{f(t/\tau)}{\sqrt{\tau}} \,.
\ee
Then, Equation~\eqref{eqn:QEIminflat} becomes
\vspace{-6pt}{}
\be
\label{eqn:QEIminflatscale}
\frac{1}{\tau} \int_\gamma dt\, \langle \nord{T_{\mu \nu} t^\mu t^\nu} \rangle_{\psi}\  f(t/\tau)^2 \,  \geq -\frac{C}{\tau^4} \,,
\ee
where the constant $C$ depends on the function $f$. For~short smearing times $\tau$, the lower bound approaches negative infinity. For~$\tau \to \infty$, the bound goes to~zero.

The first difference QEI for general spacetime curvature was derived by Fewster~\cite{Fewster:1999gj}. Difference QEIs on curved spacetimes have the disadvantage in that there is no preferred reference state. A~few years later, Fewster and Smith~\cite{Fewster:2007rh} derived the first absolute QEI for general curved spacetimes. The~QEI takes the form
\vspace{-6pt}{}
\bea
\label{eqn:generalQEI}
&&\int_\Sigma d\text{vol}(x) f^2(x) \coin{ \mathcal{D} \otimes \mathcal{D}(W_\psi-H_{(k)})} \geq \nonumber\\ && \qquad \qquad -2\int_{\mc D} \frac{d^n\xi}{(2\pi)^n} \left[ |h_\kappa|^{1/4} f_\kappa \otimes |h_\kappa|^{1/4} f_\kappa \vartheta_\kappa^* (\mc D \otimes \mc D \tilde{H}_{(k)}) \right]^\wedge (-\xi,\xi) \,,
\eea
where $\mc D$ is a partial differential operator of order zero or one, with~smooth real-valued coefficients. The~notation $[\cdot]^{\wedge}(\xi_1, \xi_2)$ means a bi-Fourier transform in two arguments $x$ and $x'$. $\Sigma$ is a \emph{small sampling domain} \cite{Fewster:2007rh, Fewster:2018pey}, which is an~open subset of $(\mathcal{M},g)$ that (i) is contained in a globally hyperbolic convex normal neighbourhood of $M$ and (ii) may be covered by a single hyperbolic coordinate chart. Additionally, it is required that the coordinate speed of light is bounded. We may express the hyperbolic chart using a map $\kappa$. Any function $f$ on $\Sigma$ determines a function $f_\kappa = f \circ \kappa^{-1}$ on $\Sigma_\kappa=\kappa(\Sigma)$. We have the pullback $\vartheta: \Sigma \times \Sigma \to \mathcal{M} \times \mathcal{M}$. Here, $h$ is a Lorentzian metric on $\Sigma$, and $h_\kappa$ is the determinant of the matrix $\kappa^*h$.

The main idea of the proof for both general and absolute inequalities is the following: we note that the left-hand side can be written as 
\vspace{-6pt}{}
\be
\label{eqn:differencegeb}
\langle \nord{\rho} (f)\rangle_\psi=\langle \rho^{\text{ren}} (f)\rangle_\psi-\langle \rho^{\text{ren}} (f)\rangle_{\psi_0} \,,
\ee
for the difference inequalities. For~absolute ones, the second part can be replaced by the appropriate derivatives of the Hadamard parametrix. Now, the challenge is to show that the first term of Equation~\eqref{eqn:differencegeb} is positive, while the second is finite. This is done by utilizing the fact that, for free fields, the operator $\Tsplit$ is of the positive type and~symmetric. 

This is not true, for~example, for~the $\Tsplit$ operator of nonminimally coupled fields, so these inequalities cannot be used in this case. State-dependent bounds have been derived for the nonminimally coupled field~\cite{Fewster:2007ec, Fewster:2018pey, Fliss:2023rzi}. 

The general inequality of Equation~\eqref{eqn:generalQEI} can be used to derive the explicit dependence of the bound on spacetime curvature. This was done in Ref.~\cite{Kontou:2014tha}, where a worldline QEI for the energy density was derived for the minimally coupled free massless scalar. The~spacetime curvature considered was small \endnote{In the sense that only terms of first order in curvature were considered.}. The~bound takes the form

\begin{adjustwidth}{-\extralength}{0cm}
\centering 
{\small
\bea
\label{eqn:curvbound}
\int_{\gamma} dt \,f(t)^2\langle T^{ren}_{\mu \nu} t^\mu t^\nu \rangle_{\psi}
(t,0) \geq- \frac{1}{16\pi^2} \bigg\{&&I_1+\frac{5}{6}\Rmax~J_2\\
&+&\Rmax''\left[\frac{23}{30}J_3+\left(\frac{43}{40}+16\pi^2(24|a|+11|b|)\right) J_4\right]
\nonumber\\
&+&\Rmax''' \left[\frac{163\pi+14}{96\pi}J_5
+\frac{7(2\pi+1)}{192\pi}(4J_6+J_7)\right] \bigg\}\,.\nonumber
\eea}
\end{adjustwidth}

Here,  the $J_i$s are integrals over the test function $f$, and the constants $a$ and $b$ express the ambiguities in the renormalization of the stress tensor in a general curved background. The~curvature is considered bounded in the following sense:
\be 
\label{Rmax}
|R_{\mu \nu}| \leq \Rmax \qquad |R_{\mu \nu,\sigma \rho}| \leq \Rmax'' \qquad |R_{\mu \nu,\sigma \rho \lambda}| \leq \Rmax'''\,.
\ee
The coordinate system used is modified Fermi coordinates described in detail in~\cite{Kontou:2012kx,Kontou:2012ve}. It is interesting to note that no other components of the Riemann tensor appear in the bound at first~order. 

These examples do not cover all the progress made in deriving QEIs. For~more extensive reviews, see~\cite{Kontou:2020bta,Fewster2017QEIs}. We should mention the results on Maxwell, Proca~\cite{Fewster:2003ey, Pfenning:2001wx}, and Dirac~\cite{Fewster:2003zn, Dawson:2006py, Smith:2007pp} fields. In~terms of interacting fields, there are no QEIs to date for four-dimensional spacetimes. However, there are results for two-dimensional CTFs~\cite{Fewster:2004nj} and two-dimensional integrable interacting QFTs~\cite{Bostelmann:2013mxa, Frob:2022hdj, Bostelmann:2023ety, Cadamuro:2024lzc}.

\subsection{Null Quantum Energy~Inequalities}
\label{sub:nullqei}

In the previous section, I did not mention any results for QEIs averaged over a null geodesic segment. These types of inequalities are particularly important for constraining wormhole solutions, as seen in the next~section. 

First, we look at results in two-dimensional spacetimes. The~first such result was by Ford and Roman~\cite{Ford:1994bj}. It was a null averaged QEI for the massless Klein--Gordon field in two-dimensional Minkowski space using a Lorentzian sampling function. The~result was generalized to general functions~\cite{Flanagan:1997gn} and any spacetime conformal to Minkowski~\cite{Flanagan:2002bd} by~Flanagan.

Fewster and Hollands~\cite{Fewster:2004nj} derived a general QEI for two-dimensional CFTs on Minkowski for timelike and null geodesics. The~CFTs considered are the unitary, positive energy conformal field theories with a stress--energy tensor. For~those, and~for averaging over null geodesics, the QEI is
\vspace{-6pt}{}
\be
\label{eqn:conf2d}
\int_\gamma d\lambda \, f(\lambda)^2 \langle \Tren_{\mu \nu} \ell^\mu \ell^\nu \rangle_{\psi}  \geq - \frac{c}{48\pi}  \int  \frac{\left( f'(\lambda) \right)^2}{f(\lambda)} d\lambda \,,
\ee
where $c$ is the central charge of the theory. The~bound also holds for all spacetimes globally conformal to Minkowski~\cite{Freivogel:2020hiz}.

It is worth mentioning that the QEI bounds by Flanagan~\cite{Flanagan:1997gn, Flanagan:2002bd} for the Klein--Gordon field and by Fewster and Hollands~\cite{Fewster:2004nj} for CFTs are sharp. This has not been proven for any of the four-dimensional QEIs bounds~discussed. 

An important question is whether or not we can have a finite lower bound over a finite null geodesic segment as we have in the timelike case for more than two dimensions. In~other words, is 
\be
\int_\gamma d\lambda f(\lambda)^2 \langle T^{ren}_{\mu \nu} \ell^\mu \ell^\nu \rangle_{\psi} \,,
\ee 
where $\gamma$ is a null geodesic, bounded from below over the class of Hadamard states $\psi$? The answer to that question, at~least for the minimally coupled free scalar field on Minkowski, is ``no'', as Fewster and Roman~\cite{Fewster:2002ne} in 2002 found a Hadamard state with no lower bound. Finding one state, a~counter-example, is sufficient to show that such a lower bound does not~exist. 

In their analysis, they used states that are superpositions of the vacuum and two-particle states. In~these states, the~excited modes are those whose three momenta lie in a cone centered around the null vector that the stress tensor is contracted with. In~the limit that the three momenta become arbitrarily large, the~null energy density averaged over a null geodesic segment can become arbitrarily~negative. 

It is interesting to consider whether or not there is a deeper reason for the nonexistence of a lower bound. A~remark of Buchholz mentioned in~\cite{Fewster:2002ne} suggests that this is due to the fact that, in~any algebraic quantum field theory in Minkowski space of dimension $n>2$ subject to reasonable conditions, there are no nontrivial observables localized on any bounded null line segment. This could perhaps provide an argument for the nonexistence of null QEIs, even for more general field theories, but~such a result does not yet~exist. 

In the same work, Fewster and Roman~\cite{Fewster:2002ne} proved that the average of the renormalized null energy over a timelike geodesic is indeed finite and the corresponding QEI takes \mbox{the form} 
\be
\label{eqn:nulltimelike}
\int_\gamma dt f(t)^2 \langle T^{ren}_{\mu \nu} \ell^\mu \ell^\nu \rangle_\psi \geq -\frac{(\ell_\mu t^\mu)^2}{12\pi^2} \int_\gamma dt (f''(t))^2 \,,
\ee
where $t^\mu$ is the tangent vector of the timelike~geodesic. 

As the divergence of the average null energy in the Fewster--Roman counterexample happens when the three momenta of the excited states become unbounded, Freivogel and Krommydas~\cite{Freivogel:2018gxj} introduced an additional restriction on it. The~justification is that when we have an effective field theory, the~momenta are below the UV cutoff of the theory. The~momentum cutoff corresponds to a length cutoff $\ell_{\text{UV}}$. This leads to the following null QEI bound, 
\be
\label{eqn:SNEC}
\int_\gamma d\lambda f^2(\lambda) \langle T^{ren}_{\mu \nu} \ell^\mu \ell^\nu \rangle_{\psi} \geq - \frac{4B}{G_N} \int_\gamma d \lambda \left(f'(\lambda)\right)^2 \,,
\ee
which was conjectured in Ref.~\cite{Freivogel:2018gxj} and called the \textit{smeared null energy condition} (SNEC). This bound includes the UV cutoff as 
\be
\label{eqn:Gluv}
G_N \lessapprox \frac{\ell_{\text{UV}}^{n-2}}{N} \,,
\ee
where $N$ is the number of fields, and $n$ is the number of spacetime dimensions. The~equality holds when the UV cutoff is the minimum possible one---the~Planck length (for one field). The~SNEC was proven in Refs.~\cite{Fliss:2021gdz, Freivogel:2020hiz} for free scalars on Minkowski spacetime using an argument for the factorization of free field theories on the lightsheet. In~that case, the~constant $B$ expresses the relation between the UV cutoff and the Planck length. In~particular, the~inequality \eqref{eqn:Gluv} is saturated for $B=1/32\pi$. 

It was conjectured that the SNEC bound can hold for spacetimes with curvature and interactive fields. It is not clear if that is true, as~additional correction terms appear in curved spacetimes~\cite{Kontou:2014tha}. In~the case of interacting fields, it is unknown if such a lower bound exists. The SNEC has also been verified for an induced gravity on a brane in AdS/CFT by Leichenauer and Levine~\cite{Leichenauer:2018tnq}. 

The SNEC suffers from its dependence on the arbitrary $\ell_{\text{UV}}$ cutoff. Such a cutoff depends on the theory, and, if taken to the lowest possible value (the Planck length), it gives a very weak bound. Additionally, it is not known how to generalize the proof of the SNEC to curved spacetimes. The~existing proof relies heavily on the properties of QFT on Minkowski spacetime. Those problems can be remedied with the \textit{double smeared null energy condition} (DSNEC) \cite{Fliss:2021phs}. The~idea here is that, as averages over finite null segments diverge, one can average instead over two null directions. The~derivation of DNSEC begins with the general QEI of Equation~\eqref{eqn:generalQEI}. From~this inequality, we can calculate the bound for general curved spacetimes in a perturbative way, as was done in Ref.~\cite{Kontou:2014tha}. This work has not been done yet, but~in Ref.~\cite{Fliss:2021phs}, the DSNEC was derived for Minkowski spacetime and minimally coupled~scalars. 

The derivation includes integrating out the other spacetime directions and optimizing the bound over the domain $\mathcal{D}$, thus considering only the set of boosted domains for simplicity. The~final bound is Lorentz invariant and was derived for massive and massless scalars for an arbitrary number of dimensions. The~result in position space for massless scalars and four spacetime dimensions is
\vspace{-6pt}{}
\be
\label{eqn:DSNEC}
\int d^2x^\pm f(x^+,x^-)^2\langle \Tren_{--}\rangle_\psi \geq-\frac{16}{81\pi^2} \left(\int dx^+ (f_+''(x^+))^2\right)^{1/4} \left(\int dx^- (f_-''(x^-))^2\right)^{3/4} \,,
\ee
where $x^-$ and $x^+$ are the two null directions. The~bound assumes that the smearing function factorizes as 
\be
f(x^+,x^-)^2=(f_+(x^+))^2 (f_- (x^-))^2 \,.
\ee
There are a couple of things to note here: First, the~bound depends on the derivatives of the smearing function but not in a simple way as the SNEC. Second, it is state independent and theory independent. Ref.~\cite{Fliss:2021phs} derived the SNEC from the DSNEC by imposing the additional momentum cutoff. The~DSNEC was recently derived for nonminimally coupled fields~\cite{Fliss:2023rzi}, although this bound was found to be state dependent, as~is similarly the case for the energy density~\cite{Fewster:2007ec} and the effective energy density~\cite{Fewster:2018pey}. 

The issue of curvature should be briefly discussed for the DSNEC. Although~there is a way to generalize the DSNEC for curved spacetimes, it is unclear how only the null area where we are averaging will be modified in this case. Additionally, the~general QEI of Equation~\eqref{eqn:generalQEI} is not generally covariant. It could be brought to a covariant form, as discussed in Ref.~\cite{Fewster:2007rh}, but this requires some care in the choice of coordinates. Thus, the DSNEC of Equation~\eqref{eqn:DSNEC} currently makes sense only on Minkowski spacetime or at length scales sufficiently smaller than the curvature~scale.

\subsection{ANEC}
\label{sub:ANEC}

One might think that averaging the quantum null energy density over an entire null geodesic will have the same problems as the null QEIs described in the previous section. However, this is not the case. The~average null energy condition (ANEC) is perhaps the hardest energy condition to violate in classical and semiclassical gravity. For~quantum fields, it states that
\be
\int_\gamma d\lambda \langle T^{ren}_{\mu \nu} \ell^\mu \ell^\nu \rangle_{\psi} \geq 0 \,,
\ee
where $\gamma$ is a complete null geodesic. The~classical version is similar, with~the renormalized expectation value of the null energy density replaced by the classical null energy density. To~see how to obtain the ANEC from the DSNEC of Equation~\eqref{eqn:DSNEC}, we first rescale the smearing function so that we have dimensionless coordinates:
\vspace{-6pt}{}
\be
f(x^+,x^-)=\frac{1}{\sqrt{\delta^+\delta^-}} \mc F(x^+/\delta^+,x^-/\delta^-)=\frac{1}{\sqrt{\delta^+\delta^-}} \mc F(s^+,s^-) \,.
\ee
where $s^{\pm}=x^\pm/\delta^\pm$. Then, Equation~\eqref{eqn:DSNEC} becomes
\vspace{-6pt}{}
\bea
\label{eq:DSNECdimless}
&&\int \frac{d^2x^{\pm}}{\delta^+ \delta^-} \mc F(s^+,s^-)^2\langle \Tren_{--}\rangle_{\psi}\geq \nonumber \\
&& \qquad \qquad -\frac{16}{81\pi^2} \frac{1}{\delta^+ (\delta^-)^3} \left(\int ds^+ (\mc F_+''(s^+))^2\right)^{1/4} \left(\int ds^- (\mc F_-''(s^-))^2\right)^{3/4} \,. 
\eea
To obtain the ANEC, we take the limits $\delta^+ \to 0$ and $\delta^- \to \infty$ while holding $\delta^+\delta^-\equiv\alpha^2$ fixed. Then, we require that the smearing function satisfies
\vspace{-6pt}{}
\be
\label{eqn:condfun}
\lim_{x^+ \to 0} \lim_{x^- \to \infty} \frac{1}{\delta^+} \mc F(x^+/\delta^+,x^-/\delta^-)^2=A \delta(x^+-\beta) \,,
\ee
where $A$ and $\beta$ are real numbers. Then, from Equation~\eqref{eq:DSNECdimless}, we have
\vspace{-6pt}{}
\be
\int dx^- \langle \Tren_{--}\rangle_{\psi}\geq- \lim_{\delta^+ \to 0} -\frac{16}{81\pi^2} \frac{\delta^+}{A\alpha^4} \left(\int ds^+ (\mc F_+''(s^+))^2\right)^{1/4} \left(\int ds^- (\mc F_-''(s^-))^2\right)^{3/4}=0 \,,
\ee
thus recovering~the ANEC.

This proof is not the most general one of the ANEC. In~fact, the ANEC has been proven in the context of semiclassical gravity in multiple ways and for a variety of situations. Here, we will only mention some of the existing~proofs. 

Yurtsever, who proved that the ANEC holds for conformally coupled scalar fields on asymptotically flat two-dimensional spacetimes~\cite{Yurtsever:1990gx}, was also the first to suggest that the ANEC holds generally. Wald and Yurtsever~\cite{Wald:1991xn} proved that the ANEC holds for all Hadamard states of a massless scalar field in any globally hyperbolic, two-dimensional spacetime along any complete, achronal null geodesic. Achronal null geodesics are those for which there are no two points that can be connected by a timelike path. As~we will see, these geodesics are important in the study of wormholes. Flanagan and Wald~\cite{Flanagan:1996gw} studied spacetimes perturbatively close to Minkowski and provided the first proof of the ANEC incorporating~backreaction. 

For general quantum fields in two-dimensional Minkowski space that have a mass gap, the~ANEC was proven by Verch~\cite{Verch:1999nt}. For~conformal field theories in two-dimensional Minkowski space, the~ANEC was established by Fewster and Hollands~\cite{Fewster:2004nj}, which was derived by the QEI of Equation~\eqref{eqn:conf2d}.

There exist a variety of holographic proofs of the ANEC assuming the AdS/CFT correspondence. Selectively, Ref.~\cite{Kelly:2014mra} proved the ANEC for CFTs in Minkowski space. The~ANEC has also been proven from the quantum null energy condition~\cite{Bousso:2015wca}, which is a~local condition that bounds the renormalized null energy using the derivatives of von Neumann entropy. Faulkner~et~al.~\cite{Faulkner:2016mzt} have argued that the ANEC can be derived directly using modular Hamiltonians for a range of interacting QFTs on Minkowski~spacetime. 

In terms of general curved spacetimes, a~proof by Fewster, Olum, and Pfenning~\cite{Fewster:2006uf} derived the ANEC from a QEI for a flat spacetime with boundaries for~geodesics which stay a finite distance away from the boundary. Using a variation of this argument, Kontou and Olum~\cite{Kontou:2015yha} proved the achronal ANEC in spacetimes with small curvature using the QEI of Equation \eqref{eqn:curvbound}. 

What about ANEC violations? So far, there are three kinds of violations~shown:
\begin{enumerate}
\item 
Violations of the ANEC on chronal null geodesics.
\item 
Violations of the non-self consistent ANEC, meaning that it does not satisfy the (classical or semiclassical) Einstein equation.
\item 
Violations of the order of Planck scale.
\end{enumerate}

For the first category of violations, the~simplest example is a quantum scalar field in a Schwarzschild spacetime around a black hole~\cite{Visser:1996iv, Levi:2016quh}. Similarly, the chronal ANEC is violated in a spacetime with a compactifed spatial dimension~\cite{Klinkhammer:1991ki}.

For the second category, a~simple example is that of a conformal transformation, as pointed out in Ref.~\cite{Wald:1991xn} and expanded by Visser~\cite{Visser:1994jb}. Consider a conformal transformation of the metric
\be
\tilde{g}_{\mu \nu}=\Omega^2 g_{\mu \nu} \,,
\ee
where $\Omega^2$ is the conformal factor. Now, the stress tensor can transform as 
\vspace{-6pt}{}
\be
\tilde{T}^{\mu}_{\, \, \, \nu}=\Omega^{-4} (T^{\mu}_{\, \, \, \nu}-8 \alpha Z^\mu_{\,\,\, \nu} \ln{\Omega}) \,,
\ee
which is a special case of the general result derived in~\cite{Page:1982fm}. Here, $\alpha$ depends on the spin of the field and 
\be
Z^\mu_{\,\,\, \nu}=\left( \nabla_\rho \nabla^\sigma +\frac{1}{2} R^{\,\,\,\sigma}_{\rho} \right) C^{\rho \mu}_{\,\,\,\,\,\, \sigma \nu} \,,
\ee
where $C^{\rho \mu}_{\,\,\,\,\,\, \sigma \nu}$ is the Weyl tensor. This is often called \textit{anomalous scaling} of the stress tensor. The~ANEC integral in the conformal spacetime is
\vspace{-6pt}{}
\be
\int_\gamma d\lambda \, \tilde{T}^{\mu}_{\, \, \, \nu} \ell_\mu \ell^\nu=\Omega^{-4} \left(\int_\gamma d\lambda \, T^{\mu}_{\, \, \, \nu} \ell_\mu \ell^\nu-8\alpha \ln{\Omega }\int_\gamma d\lambda \, Z^\mu_{\,\,\, \nu} \ell_\mu \ell^\nu \right)\,.
\ee
Assuming that ANEC was obeyed in the original spacetime, we need 
\vspace{-6pt}{}
\be
\label{eqn:confviol}
8\alpha \ln{\Omega }\int_\gamma d\lambda \, Z^\mu_{\,\,\, \nu} \ell_\mu \ell^\nu > \int_\gamma d\lambda \, T^{\mu}_{\, \, \, \nu} \ell_\mu \ell^\nu \,,
\ee
to have a violation in the conformal spacetime. Now, let us assume that the Einstein equation is obeyed in the original spacetime so that the integrals of  Equation~\eqref{eqn:confviol} are comparable in magnitude. Then, for~the equation to be obeyed, we need $8\alpha |\ln{\Omega}| \gtrapprox 1$, and for scalar fields with $\alpha=(1920 \pi^2)^{-1}$, we have $\Omega \gtrapprox e^{240 \pi^2}$ or $\Omega \lessapprox e^{-240 \pi^2}$. As~was first observed in~\cite{Graham:2007va}, this is either an enormous dilation or an enormous contraction that is not compatible with self-consistent semiclassical gravity. The~reason is that, in this case, the~Einstein tensor scales with $\Omega^{-2}$, while the stress--energy tensor scales with $\Omega^{-4}$. Thus, consistency cannot be imposed in either~case. 

I should briefly comment on the notion of self-consistency. The~solution of the semiclassical Einstein equation is complex and requires a state and a metric that simultaneously solve the system of equations. The~existence of solutions has been shown only for spacetimes with a high level of symmetry, for~example, cosmological~\cite{Meda:2020smb, Gottschalk:2018kqt} or static~\cite{Sanders:2020osl}. Sometimes, a~classical background metric is assumed, which is consistent with classical matter that usually obeys the NEC. In~the context of the ANEC, it is mostly used to show the nonself consistent character of some counterexamples, such as the one presented, and not as a derivation~method.

\textls[-15]{Finally, there are violations that occur if the geodesic is not integrated in the transverse direction over at least a few Planck lengths. Examples are included in Ref.~\cite{Graham:2007va}, where violations in spacetimes conformal to Minkowski were considered. Those can be eliminated either by averaging over transverse directions or imposing self-consistency. The~proof of Flanagan and Wald~\cite{Flanagan:1996gw} also allows for violations if no averaging in the transversal directions is included. These violations are in a range of validity outside the scope of semiclassical~gravity. }

The self-consistent achronal ANEC is free of violations in the realm of semiclassical gravity. Its prominent role was emphasized in Ref.~\cite{Graham:2007va} and also discussed in Ref.~\cite{Kontou:2020bta}. In~Section~\ref{sec:short}, I will demonstrate 
 its importance for~wormholes.

\section{Wormhole~Basics}
\label{sec:wormholes}
\unskip

\subsection{Description and~History}

\begin{figure}[H]
	\includegraphics[height=8cm]{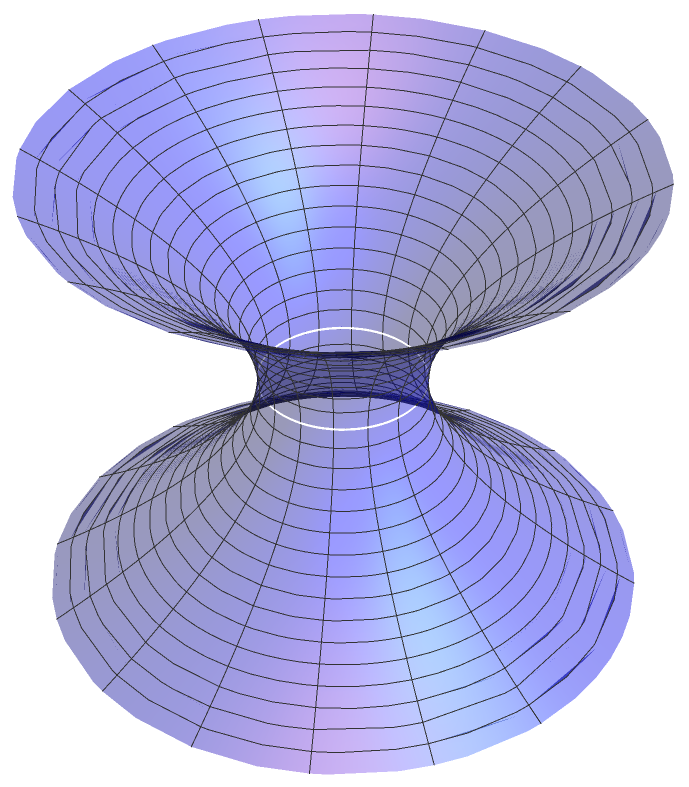}
	\caption{Diagram of a generic wormhole in two spatial dimensions embedded in three spatial dimensions. The~``mouths'' and ``throat'' are~visible.}
	\label{fig:wormholegen}
\end{figure}

A wormhole is generically a bridge that connects two parts of spacetime. It is connecting a ``wrapping'' or folding of spacetime, and it involves the existence of nontrivial topologies. A~diagram of a generic wormhole is presented in Figure~\ref{fig:wormholegen}. The~``mouths'' are the openings of the wormhole, usually surrounded by asymptotically flat regions, while the ``throat'' is the bridge connecting the two~regions.

We can categorize them in interuniverse wormholes: those that connect parts of our universe and intrauniverse wormholes and those that connect ``our'' universe with ``another'' universe~\cite{Visser:1995cc}. The~concept of the multiverse and the different categories of universes is a complex and debatable one, and it goes beyond the scope of this paper. When we are referring to generic wormholes, we will assume they are intrauniverse ones. We can further categorize wormholes to ``traversable'' ones that allow the passage of a macroscopic object through their throat. One could add as a requirement that travel should take a finite time, and if the traveler is a living being, they arrive healthy on the other side. Any other kind of wormhole is referred to as ``nontraversable''. Any wormhole containing curvature singularities is probably nontraversable, even if the singularity is not on the throat. The~reason is the strong tidal forces near the singularity, which can only be avoided if the singularity is sufficiently far from the trajectory of the observer.  Wormholes that contain event horizons are also considered nontraversable by some, as it takes infinite time to get to the horizon from the point of view of an external observer~\cite{Visser:1995cc}.

A `short' wormhole is one where the proper time of the observer travelling through it is shorter that the travel in ambient space (outside the wormhole). In~a `long' wormhole, travelling between the mouths on the outside is shorter in proper time. `Short' wormholes are obviously more interesting. However, as~we are going to see, they allow for causality violations, and they are subject to stricter~constraints. 

The work of Flamm~\cite{flamm1916beitrage, flamm2015republication} conducted very early in the history of general relativity is cited as the first related to wormhole geometries. However, the~first actual wormhole solution was constructed by Einstein and Rosen in 1935~\cite{Einstein:1935tc}. Their motivation was not to create a wormhole, but~rather to construct a singularity-free particle model. They considered the isometric embedding of the equatorial section of the Schwarzschild solution to a \mbox{three-dimensional} Euclidean space with two flat sheets, and the ``particle'' was represented as a bridge between these sheets. The~Einstein--Rosen wormhole is not a traversable one, as an observer inevitably ends up on the Schwarzschild curvature~singularity.

Wheeler~\cite{Wheeler:1955zz} presented the next contribution to the field. He wanted to study ``geons'', which are unstable solutions to the coupled Einstein--Maxwell field equations. In~his work appears a notion of what we today call a wormhole: ``\ldots a metric which on the whole is nearly flat except in two widely separated regions, where a double-connectedness comes into evidence \ldots'' \cite{Wheeler:1955zz}. 

The Kerr solution of the vacuum Einstein equation provides a different wormhole-like scenario.  The~solution was discovered in 1963, and it generally describes a rotating black hole with mass $M$ and angular momentum $L$. The~case of $a=L/M<M$ is what is considered to describe a physical black hole. However, the~maximally extended solution with $a>M$ possesses no horizon~\cite{Visser:1995cc}. Instead, it has a naked (ring) singularity between two connected asymptotically flat~regions. 

An important breakthrough in the field came with the work of Morris, Thorne, and~Yurtsever~\cite{Morris:1988tu}. They provided a mechanism to keep the wormhole traversable and showed how it can be used as a time machine. This will be discussed in detail in Section~\ref{sub:time}. 

A significant amount of work has been done in nonspherically symmetric wormholes using the thin shell approximation. In~that, a~thin shell of matter generates a gravitational field, where the metric is continuous, the~Christoffel connection has a step function discontinuity, and the Riemann tensor has a delta function contribution. The~formalism was first introduced by Sen~\cite{sen1924grenzbedingungen} and Lanczos~\cite{lanczos1924flachenhafte} and developed in more recent work~\cite{Barrabes:1991ng}. A~detailed description of the formalism can be found in Ref.~\cite{Visser:1996iv}.

Wormholes also appear in the context of the AdS/CFT correspondence. The~most famous example is the wormhole of Gao, Jafferis, and Wall~\cite{Gao:2016bin}. Starting with an eternal AdS black hole, the~wormhole is an Einstein--Rosen bridge that, as~discussed, is nontraversable. However, in~the context of AdS/CFT correspondence, the geometry is dual to two CFTs. Adding a coupling between the CFTs adds a nonlocal coupling between the two asymptotic regions of the gravity part. This coupling allows the wormhole to become traversable. Though~is has been shown~\cite{Freivogel:2019whb} that the wormhole remains open only for a time scale shorter than the Planck time, some information quanta might be able to be~transmitted.

We should briefly comment on the concept of Euclidean wormholes. All the previous wormholes mentioned are Lorentzian, meaning that they describe the physical spacetime. Moving to the Euclidean signature, wormhole constructions are mostly unrestricted from energy conditions or most topology constraints. These wormholes are usually thought of as gravitational instantons~\cite{Giddings:1987cg} and not as spacetime bridges. They are of great importance for particle physics, starting from the work of Coleman~\cite{Coleman:1988tj}.

\subsection{Violations of Energy~Conditions}

Since the 1970s, it became apparent that the kind of matter required to create and support traversable wormholes is one that violates energy~conditions.

Tipler~\cite{Tipler:1977eb} first proved that the creation of a wormhole required the violation of the WEC. The~starting point is work by Geroch~\cite{Geroch:1967fs},  who used the notion of \textit{causality}, meaning that a continuous choice of the forward light cone can be made, and that no closed timelike curves exist. If~causality is maintained, topology changes in a compact manifold are forbidden. Tipler replaced causality using the WEC and Einstein’s equation. Additionally, he showed that if topology changes happen in a finite region, the~change must be accompanied with~singularities. 

The maintenance of a traversable wormhole means that a two-sphere surrounding one mouth, as~seen through the wormhole from the other mouth, is an outer trapped surface. A~proposition in Ref.~\cite{hawking2023large} adapted by~\cite{Morris:1988tu} implies that the wormhole's stress--energy tensor must violate the average WEC (AWEC). 

A weaker energy condition, the~ANEC, was used to constrain topology changes by Friedman, Schleich, and Witt~\cite{Friedman:1993ty}. In~particular, they proved the following~theorem.

\begin{Theorem}
	\label{the:topological}
Let $\gamma_0$ be a timelike curve with a past endpoint in $\mathcal{J}^-$ and a future endpoint in $\mathcal{J}^+$ that lies in a simply connected neighborhood of $\mathcal{J}$. If~it is 
 asymptotically flat, and globally hyperbolic spacetime satisfies the averaged null energy condition, then every causal curve from $\mathcal{J}^-$ to $\mathcal{J}^+$ is deformable to $\gamma_0$.
\end{Theorem}

Here, $\mathcal{J}^\pm$ is the future (+) and past (-) null infinity, while $\mathcal{J}=\mathcal{J}^+ \cup \mathcal{J}^-$ is their disjoint union. This theorem prohibits topology changes in such manifolds that obey the ANEC. And~what about traversable wormhole maintenance? It seems that a similar argument to~\cite{Morris:1988tu} can be used with the ANEC. In~particular, the~divergence of null geodesics passing through the throat requires the violation of average null convergence condition, at least in cases with asymptotically flat manifolds \endnote{This argument is possible to be extendable to asymptotically AdS manifolds, but this point requires further consideration.}. The~self-consistent ANEC is thus also required to be~violated.

As discussed in Section~\ref{sub:ANEC}, the chronal ANEC can be violated in the context of semiclassical gravity. At~this point, we will split the discussion in two cases: the `short' wormholes that can lead to causality violations and the `long' wormholes that do not. The~strongest restriction to the existence of `short' wormholes is given by the achronal ANEC, discussed in Section~\ref{sub:causality}.

\section{Restrictions to `Short'~Wormholes}
\label{sec:short}

`Short' wormholes can be used to produce closed timelike curves or time machines. As~we will see, they are severely constrained by QEIs and mainly by the achronal ANEC. We begin with the description of how a wormhole can be used as a time machine and continue with the constraints provided by timelike~QEIs. 

\subsection{A Wormhole as a Time~Machine}
\label{sub:time}

Morris, Thorne, and~Yurtsever~\cite{Morris:1988tu} were the first to provide a concrete mechanism for a wormhole acting as a time machine. In~their work, they considered a wormhole described by the metric
\be
ds^2=-e^{2\Phi(l)} dt^2+dl^2+r^2(l)(d\theta^2+\sin^2{\theta} d\phi^2) \,,
\ee
introduced in~\cite{Morris:1988cz}, where $\Phi$ and $r$ are functions of the proper radial distance $l$. $l$ is zero at the middle of the throat and takes negative and positive values at each side. Starting from $l \to -\infty$ and increasing, $r$ decreases, thus reaching its minimum value $r_0$, which is the~radius of the throat. Then, it increases again reaching $r=|l|-M\ln{(|l|/r_0)}$ far from the throat. $\Phi$ remains finite everywhere, as the construction has no horizons. It is interesting to note that Ref.~\cite{Morris:1988tu} provided both a mechanism of wormhole stabilization and a speculative mechanism for construction. The~first is spherical Casimir plates on each side of the throat. Their existence provides the required negative energy to keep the throat open. For~the second, the~authors suggest the method of spontaneous wormhole production via quantum tunneling, although~no more details are~provided. 

In order to turn the wormhole into a time machine, one of the mouths has to be accelerated. The~metric for the accelerating mouth is 
\vspace{-6pt}{}
\be
\label{eqn:metric}
ds^2=-(1+glF \cos{\theta}) e^{2\Phi} dt^2+dl^2+r^2(d\theta^2+\sin^2{\theta} d\phi^2) \,,
\ee
where $g=g(t)$ is the acceleration. The~function $F=F(l)$ is a form factor that vanishes for $l\leq 0$, thus guaranteeing that only one of the mouths accelerates. Outside the left (L) [static] and right (R) [accelerating] mouths, we can define external Lorentz coordinates
\vspace{-6pt}{}
\be
ds^2 \approxeq -dT^2+dX^2+dY^2+dZ^2 \,.
\ee

The connection between Lorentz coordinates and wormhole coordinates for each mouth is given in Table~\ref{tab:wormcoord}. Note that the position of the left mouth $Z_L$ is constant, while the position of the right mouth $Z_R$ is time dependent. Here, $v(r)=dZ_R/dT_R$ is the velocity of the right mouth, and $\gamma(t)=(1-v^2)^{-1/2}$. The~right mouth's maximum acceleration $g_{\max}$ and distance of the two mouths $S$ satisfy
\be
g_{\max} S \ll 1 \,, \qquad g_{\max} \left| \frac{dg}{dt}\right|^{-1} \gg S \,.
\ee
Then, the shape and size of the wormhole are kept nearly~constant. 

\begin{table}[H]
\caption{Connection between outside Lorentz coordinates and wormhole coordinates for both~mouths.}
	\label{tab:wormcoord}
	\centering
	\begin{tabularx}{\textwidth}{CC} \toprule 
		\textbf{Right Mouth} & \textbf{Left Mouth} \\ \midrule 
		$T=T_R+v\gamma l \cos{\theta}$ & $T=t$ \\ \midrule 
		$Z=Z_R+\gamma l \cos{\theta}$ &  $Z=Z_L+l \cos{\theta}$ \\ \midrule 
		$X=l\sin{\theta}\cos{\phi}$&  $X=l\sin{\theta}\cos{\phi}$\\ \midrule 
		$Y=l\sin{\theta}\sin{\phi}$& $Y=l\sin{\theta}\sin{\phi}$ \\ \bottomrule
	\end{tabularx}
\end{table}

\begin{figure}[H]
	\includegraphics[height=8cm]{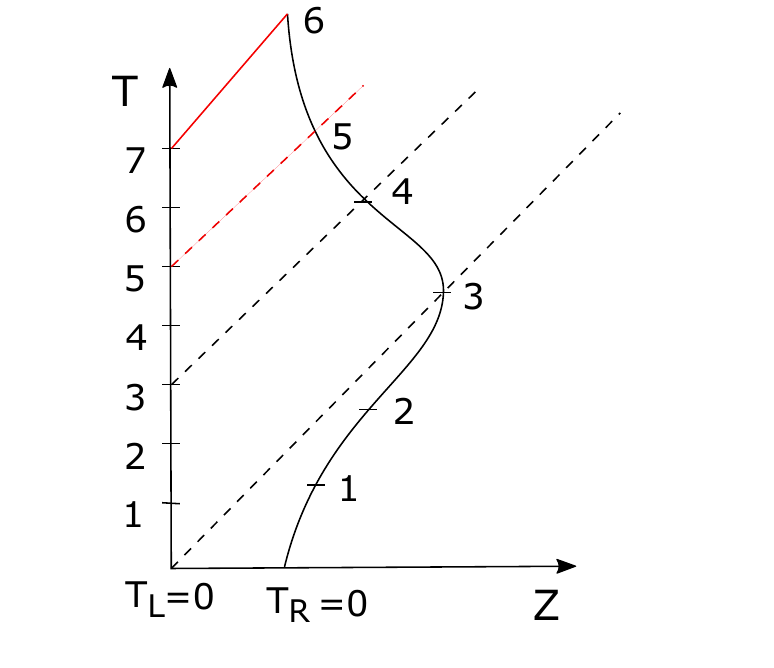}
	\caption{Diagram
 of the wormhole time machine described in Ref.~\cite{Morris:1988tu}. The~right mouth of the wormhole is accelerated and thus it has shorter proper time. After~$T_L=5$, closed timelike curves~exist. }
	\label{fig:timemachine}
\end{figure}

Figure~\ref{fig:timemachine} shows how a wormhole with one accelerating mouth acts as a time machine. The~two mouths begin at $T_R=T_L=0$. The~the right mouth begins accelerating. The~proper time of the accelerating mouth is shorter than the static one. At~some point ($T=5$ in Figure~\ref{fig:timemachine}), the lightlike path outside of the wormhole connects the mouths at equal proper times. After~that is the region of closed timelike curves and causality violation. Ref.~\cite{Morris:1988tu} proved that the Cauchy horizon $H^+(S)$, that is, the boundary of the closed timelike curves region, is immune to instabilities.To see how that works, imagine an observer at $T_L=7$ that moves along the red timelike path reaching the right mouth. But,~the right mouth is at $T_R=6$. So, when the observer jumps in the wormhole, they reach the right mouth at an earlier time than the one they departed. Note that this time machine allows travel back in time only up to the point that it was created. In~this example, the mouths are considered very close together in wormhole coordinates. But,~the same method works for any distance that is shorter than the outside~space. 
\begin{figure}
	\caption*{The dashed lines show the lightlike paths between the mouths. The red dashed line is the lightlike path that connects them at equal proper times. After that the right mouth is at earlier time (red line) allowing for traveling back in time.}
\end{figure}

In Ref.~\cite{Morris:1988tu}, it was shown that the AWEC needs to be violated in this wormhole construction. However, as~mentioned, the~AWEC can be violated in the context of QFT in a semiclassical setting. So, when it was described, the~ Morris, Thorne, and Yurtsever wormhole was seemingly plausible. As~we will see, QEIs and the achronal ANEC provide stronger~constraints.

\subsection{Timelike QEI~Constraints}

In this section, I review constraints on known, possibly short wormhole solutions from timelike QEIs. I follow Refs.~\cite{Ford:1995wg, Fewster:2005gp}. The~wormhole metric I consider is the one of Equation~\eqref{eqn:metric} first introduced by Morris and Thorne~\cite{Morris:1988cz}. In~terms of the radial coordinate $r$, the~metric can be written as
\vspace{-6pt}{}
\be
ds^2=-e^{2\Phi(r)} dt^2+\frac{dr^2}{(1-b(r)/r)}+r^2(d\theta^2+\sin^2{\theta} d\phi^2) \,.
\ee
The point $r=r_0$ corresponds to the minimum of $r$ at the throat. Here, $b(r)$ is the ``shape function'' of the wormhole, and $\Phi(r)$ is called the ``red-shift function'' defined for $r\geq r_0$. The~proper radial distance $l$ in terms of $b(r)$ is
\vspace{-6pt}{}
\be
\label{eqn:bandell}
l=\int_{r_0}^{r(l)} \frac{dr'}{(1-b(r')/r')^{1/2}} \,.
\ee
Using the Einstein equation with $G_N=1$,
\vspace{-6pt}{}
\be
G_{\mu \nu}=8\pi T_{\mu \nu} \,,
\ee
we obtain the components of the stress--energy tensor. Working in a static orthonormal frame $\hat{e}_i$, we pick a null vector $\ell^\mu= \hat{e}_t-\hat{e}_r$. Then, the NEC is 
\vspace{-6pt}{}
\be
\label{eqn:necforr}
T_{\mu \nu} \ell^\mu \ell^\nu=-\frac{e^{2\Phi}}{8\pi r} \frac{d}{dr} \left( e^{-2\Phi} \left( 1-\frac{b}{r} \right) \right) \,,
\ee
which---at the minimum of the throat---becomes 
\vspace{-6pt}{}
\be
\label{eqn:necmin}
T_{\mu \nu} \ell^\mu \ell^\mu=\frac{b_0'-1}{8\pi r_0^2} \,,
\ee
where $b'_0=b'(r_0)$. Using Equation~\eqref{eqn:bandell}, we have that 
\vspace{-6pt}{}
\be
b'_0=1-2r_0 \frac{d^2r}{d\ell^2}\bigg|_{\ell=0} \,.
\ee
Since $r$ has a minimum in the throat, the~second derivative is non-negative, and thus, $b_0 \leq 1$. So, the NEC is violated unless $b'_0=1$. Then, using Equation~\eqref{eqn:necforr}, we notice that it vanishes at $r_0$, but~it is strictly positive for any $r>r_0$ using the mean value theorem. It follows that there is a point at which the NEC is violated~\cite{Fewster:2005gp}.

The timelike QEIs used in Refs.~\cite{Ford:1995wg, Fewster:2005gp} to provide constraints are the bound on the energy density \eqref{eqn:QEIminflat} and the bound on the null energy \eqref{eqn:nulltimelike}. Both of them are Minkowski spacetime bounds. These can be used on curved spacetimes as long as $\tau \ell_{\min} \ll 1$, where $\ell_{\min}$ is the smallest radius of curvature. This assumption became exact and was proven in Ref.~\cite{Kontou:2014tha}, where a QEI for the energy density was derived for spacetimes with general but small curvature, as discussed in Section~\ref{sub:QEIs}. 

Using the timelike QEI for the null energy \eqref{eqn:nulltimelike} and following Ref.~\cite{Fewster:2005gp}, we assume that a state $\psi$, for~a time $t$ less than the support of $f$, has 
\vspace{-6pt}{}
\be
\label{eqn:epsilonbound}
\langle T^{ren}_{\mu \nu} \ell^\mu \ell^\nu \rangle_\psi \leq \mathcal{E} \,,
\ee
where $\mathcal{E}$ is a real constant. Then,
\vspace{-6pt}{}
\be
\label{eqn:epsilondef}
\mathcal{E} \geq -\frac{(\ell_\mu U^\mu)^2}{12\pi^2} \frac{\int_\gamma dt (f''(t))^2}{\int_\gamma dt f(t)^2} \,.
\ee
Now, the task is to find a function $f$ that gives the maximum negative bound to Equation~\eqref{eqn:epsilondef}.  We can reformulate this problem as finding the lowest eigenvalue of the operator~\cite{Fewster:1999kr}
\vspace{-6pt}{}
\be
L=\frac{d^4}{dt^4} \,.
\ee
So, we have to solve $f''''=\lambda f$ subject to the initial conditions $f=f'=0$ for the boundary of the interval. The~solution was found in Ref.~\cite{Fewster:2006kt} for an interval with length $\tau$, and it gives
\vspace{-6pt}{}
\be
\label{eqn:qeiepsilon}
\mathcal{E} \geq -\frac{C(\ell_\mu U^\mu)^2}{\tau^4} \,,
\ee
with $C\approx 4.23$. Then, for any Hadamard state $\psi$ obeying \eqref{eqn:epsilonbound} for $\tau \ll \ell_{\min}$, then $\mathcal{E}$ and its duration $\tau$ must obey \eqref{eqn:qeiepsilon}.

For a static trajectory and using Equation~\eqref{eqn:necmin}, we have 
\vspace{-6pt}{}
\be
\mathcal{E}=\frac{b_0'-1}{8\pi r_0^2 l_{\text{pl}}} \geq -\frac{C}{\tau^4} \,.
\ee
Now, since $\tau \ll \ell_{\min}$,  we assume that $\tau=10^{-2} \ell_{\min}$. Next, we have to estimate $\ell_{\min}$ from computation of the curvature components. The~length scales from the three nonzero curvature components are $r_0$, $r_0(1-b_0')^{1/2}$, and $(r_0/(|\Phi_0'|(1-b_0'))^{1/2}$. Here and for~simplicity, we will analyze only the first two; the third is included in Ref.~\cite{Fewster:2005gp}. The~length for traversability will be considered $\sim 10^{20} l_{\text{pl}}$, which is similar to the proton~radius.

For $\ell_{\min}=r_0$, we have
\be
r_0 \lessapprox  \frac{10^5 l_{\text{pl}}}{\sqrt{1-b_0'}} \,.
\ee
So to achieve traversability, we need $b_0' \leq 10^{-30}$, which is an~extreme fine-tuning of the wormhole. Next, for~$\ell_{\min}=r_0(1-b_0')^{1/2}$, we have
\vspace{-6pt}{}
\be
r_0 \lessapprox 10^5 l_{\text{pl}}\sqrt{1-b_0'} \,.
\ee
Then, $b_0' \leq 10^{-30}$, and $\ell_{\min} \lessapprox 10^{5} l_{\text{pl}}$. So, the minimum curvature is unphysically small in that case. The~third case leads to either an unphysically wide throat or a very fine-tuned $b_0'$.

This example shows how derived QEIs can provide constraints to wormhole solutions.  I should mention that the previous analysis was performed for a minimally coupled scalar field. An~interesting, semiclassically self-consistent wormhole solution is using nonminimal coupling to the curvature~\cite{Hochberg:1996ee}. The~throat of the wormhole derived is of Planck length or up to the order $10^2 \ell_\text{pl}$, so it is doubtful that it can be considered in the regime of the semiclassical approximation. However, it would be of interest to examine that wormhole solution using one of the QEIs for the nonminimally coupled fields derived in~\cite{Fewster:2007ec,Fliss:2023rzi} to determine its exact~validity.

\subsection{Causality Violations and the Achronal~ANEC}
\label{sub:causality}

Perhaps the stronger restriction to causality violating spacetimes was derived by Graham and Olum~\cite{Graham:2007va}. They first generalized the Theorem \ref{the:topological} by Friedman, Schleich, and Witt and then two theorems by Tipler~\cite{Tipler:1976bi} and Hawking~\cite{Hawking:1991nk} to~hold for the self-consistent achronal ANEC. Here, we present the Tipler version:

\begin{Theorem}
 An asymptotically flat spacetime $(M,g)$ cannot be geodesically complete if (a) the self-consistent achronal ANEC is satisfied, (b) the generic condition holds on $(M,g)$, (c) $(M,g)$ is partially asymptotically predictable from a partial Cauchy surface $S$, and (d) the chronology condition is violated on $J^+(S) \cap J^-(\mathcal{J}^+)$.
\end{Theorem}

$J^{\pm}(A)$ is the causal future (+) and past (-) of a set $A$. First of all, we note that this theorem only holds in the case of asymptotically flat spacetimes. Thus, it can be used for wormholes connecting two asymptotically flat regions but not other cases. Condition (b) of the theorem is the \textit{generic condition}, which (in the null version) states that every complete null geodesic contains a point where
\vspace{-6pt}{}
\be
\ell^\mu \ell^\nu \ell_{[\rho}R_{_\lambda]\mu \nu[\sigma}\ell_{\kappa]} \neq 0 \,,
\ee
where $\ell^\mu$ is its tangent vector. So, the condition states that every null geodesic is affected by matter, and thus, it is expected to hold for any realistic spacetime. Additionally, if~condition (a) holds with (b), the~complete null geodesics admit conjugate points and are therefore~chronal.  

Condition (d) means there are closed timelike curves. If~there is an area with closed timelike curves, then a Cauchy horizon $H^+(S)$ exists. That horizon is composed of null generators. Ref.~\cite{Tipler:1976bi} showed that, along with condition (c), at least one generator lies on $H^+(S)$, and thus, it is complete and achronal. As~discussed, conditions (a) and (b) show that there are no achronal complete null geodesics, which leads to the~contradiction. 

Thus, there cannot be closed timelike curves in a spacetime where the self-consistent achronal ANEC holds, is generic, asymptotically flat, and partially asymptotically predictable. Given the discussion on the self-consistent achronal ANEC in Section~\ref{sub:ANEC}, solutions such as the Morris, Thorne, and~Yurtsever~\cite{Morris:1988tu, Morris:1988cz} wormhole seem to be ruled out in semiclassical~gravity.

\section{Restrictions to Long~Wormholes}
\label{sec:long}

In this section, I examine restrictions from QEIs to `long' wormholes. In~the previous section, I described how the achronal ANEC is sufficient to rule out `short' wormholes. However, the~`long' ones do not have complete achronal null geodesics, so they escape the Graham and Olum~\cite{Graham:2007va} proof. As~they posses achronal null geodesic segments, null QEIs are the appropriate tool to constrain them. We will focus on the Maldacena, Milekhin, and Popov `long' wormhole~\cite{Maldacena:2018gjk} and examine it under the SNEC and DSNEC~bounds. 

\subsection{The Wormhole of Maldacena, Milekhin, and~Popov}

\textls[-15]{Perhaps the most famous `long' wormhole is that of Maldacena, Milekhin, and Popov~\cite{Maldacena:2018gjk}. Its advantage is that it only uses matter predicted from the standard model and not any speculative particles. In~particular, it is a solution of an Einstein--Maxwell theory with charged massless fermions, which give rise to negative energy required to keep the throat open. As~in other wormhole solutions, only the stability of it is examined and not how one could construct~it. }

\begin{figure}[H]
	\includegraphics[height=5cm]{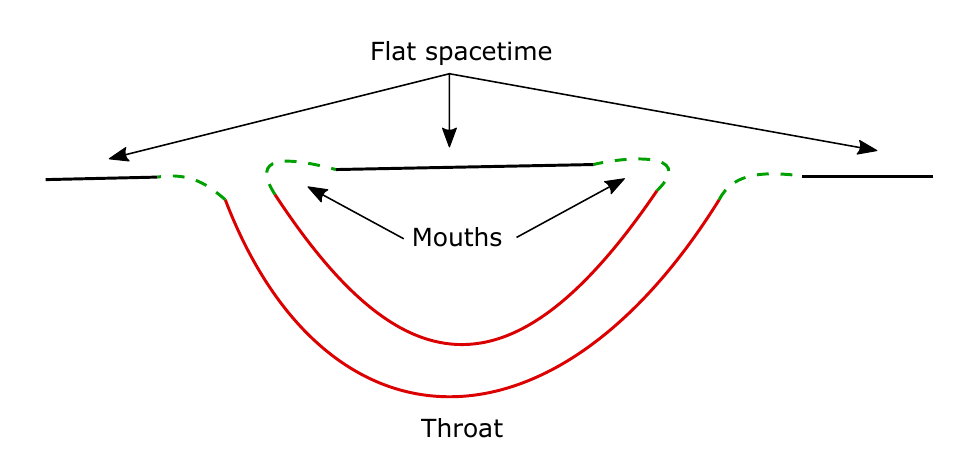}
	\caption{Schematic representation of the Maldacena, Milekhin, Popov wormhole adapted from~\cite{Maldacena:2018gjk}. It includes three regions with three different metrics that are matched: the throat (red), the~mouths (green dashed), and the region of flat spacetime (black). }
	\label{fig:mmpwormhole}
\end{figure}

The wormhole is composed of three regions: the interior or throat, the~mouths, and the flat spacetime region outside of the mouths, as shown in Figure~\ref{fig:mmpwormhole}. 
The mouths are described by the magnetically charged black hole solutions:
\vspace{-6pt}{}
\be
ds^2=-\left(1-\frac{2MG_N}{r}+\frac{r_e^2}{r^2} \right)dt^2+\left(1-\frac{2MG_N}{r}+\frac{r_e^2}{r^2}\right)^{-1}dr^2+d\Omega_2^2 \,,
\ee
where $M$ is the mass of the black hole, $\Omega_2$ is the $2$-sphere, and 
\vspace{-6pt}{}
\be
\label{eqn:rexpr}
r_e^2=\frac{\pi q^2 \ell_{\text{pl}}^2}{g^2} \,.
\ee
Here, $q$ is an integer, thus giving us the charge of the black hole, and $g$ is the coupling constant of the $U(1)$ gauge field. This geometry connects with the flat spacetime region via a transition region at $r\sim r_e$. The~two black holes are at some relative distance $d$ that is large compared to their~size.

The horizon is located at $r=r+$, where
\vspace{-6pt}{}
\be
r_+=MG_N+\sqrt{M^2G_N^2-r_e^2} \,.
\ee
We note that the near horizon geometry approaches $AdS_2 \times S_2$, which allows us to do the matching with the throat. The~wormhole interior is given by~$AdS_2 \times S_2$ geometry:
\vspace{-6pt}{}
\begin{equation}
	\label{eqn:throatmetric}
	ds^2 = r_e^2  \left[ -(\rho^2+1) d\tau^2+\frac{d\rho^2}{\rho^2+1}+d\Omega_2^2 \right] \,,
\end{equation}
The matching conditions are
\vspace{-6pt}{}
\begin{equation}
	\tau = \frac{t}{\ell}, \quad \rho = \frac{\ell(r-r_e)}{r_e^2}, \quad \mbox{with}~ 1\ll \rho, \quad \frac{r-r_e}{r_e}\ll 1, \quad 1\ll \frac{\ell}{r_e} \,,
\end{equation}
where $\ell$ is a free length scale, which is later identified as the effective length of the wormhole in two~dimensions. 

We want to apply QEIs on achronal segments of the wormhole, and~so we compute the maximum achronal segment measured in the dimensionless coordinate $\rho$. This computation was performed in Ref.~\cite{Freivogel:2020hiz}, and we summarize here. The~condition is that the length of the achronal segment has to be shorter than the rest of the geodesic inside the wormhole added to the distance between the mouths in ambient space, so
\vspace{-6pt}{}
\be
\label{eqn:archonal}
\Delta \rho \big|_{WH} <\Delta \rho\big|_{OUT} \,.
\ee
\textls[-15]{Let us  assume that the segment's middle is at $\rho=0$, and it stretches form $-\rho_0/2$ to $\rho_0/2$. Then,}
\vspace{-6pt}{}
\be
\Delta \rho \big|_{WH}=\int_{-\rho_0/2}^{\rho_0/2} \frac{d\rho}{1+\rho^2}=2\arctan{(\rho_0/2)} \,.
\ee
To find $\Delta \rho\big|_{OUT}$, we have to add the remaining pieces inside the wormhole and the distance in ambient space, which in the $\rho$ coordinate is $d/\ell+2r_e \approx d/\ell$. We have
\vspace{-6pt}{}
\be
\Delta \rho\big|_{OUT}= \int_{-\infty}^{-\rho_0/2} \frac{d\rho}{1+\rho^2}+\frac{d}{\ell}+\int_{\rho_0/2}^\infty \frac{d\rho}{1+\rho^2}=\pi-2\arctan{(\rho_0/2)}+\frac{d}{\ell} \,.
\ee
From the condition of Equation~\eqref{eqn:archonal}, the achronal segment is
\vspace{-6pt}{}
\be
\rho_0 <2\tan{\left(\frac{\pi}{4}+\frac{d}{4\ell}\right)} \,.
\ee
To proceed, we need a relationship between $d$ and $\ell$. We are interested in the regime of $d>\ell$, as we want to check the `shortest' possible wormhole in this description. However, $\pi \ell > d$, as this is the causality limit. The~maximum value of $d$ is found by minimizing the energy of the wormhole~\cite{Freivogel:2020hiz}. That numerically gives $d=\pi \ell/2.35$. Then, we have that $\rho_0< 4.13$. As~we approach the limit of a `short' wormhole $d \to \pi \ell$, then $\rho_0$ becomes larger and larger. However, the~stability of the wormhole is questionable at that~limit. 

The stress--energy tensor for $d=\pi \ell/2.35$ was computed in Ref.~\cite{Freivogel:2020hiz} using the general expression from~\cite{Maldacena:2018gjk}:
\be
\label{eqn:wormstress}
\langle T_{tt} \rangle =-\frac{q}{12 \pi^2 r_e^2 \ell^2} A \,,
\ee
where $A$ is a constant of order $1$.  We note that the negative energy comes from the two-dimensional fermions. As~expected, the~amount of negative energy increases by increasing $q$, which is the~number of fermions. In~terms of the dimensionless coordinates $\tau$ and $\rho$, we have
\vspace{-6pt}{}
\be
\label{eqn:stresstensrho}
T_{\rho \rho}=\frac{T_{\tau \tau}}{(1+\rho^2)^2}=\ell^2 \frac{T_{tt}}{(1+\rho^2)^2} \,.
\ee

\subsection{SNEC}

In the SNEC bound of Equation~\eqref{eqn:SNEC}, the null vector $\ell^\mu$ is the tangent to the geodesic parametrized with the affine parameter $\lambda$:
\vspace{-6pt}{}
\be
\int^{+\infty}_{-\infty} d\lambda f^2(\lambda) \left\langle T^{ren}_{\mu \nu} \frac{\partial x^\mu}{\partial \lambda} \frac{\partial x^\nu}{\partial \lambda} \right\rangle_{\psi} \geq - \frac{4B}{G_N} \int^{+\infty}_{-\infty} d \lambda \left(f'(\lambda)\right)^2 \,.
\ee
The coordinate $\rho$ is an affine parameter, so we can set $\lambda=\rho$:
\vspace{-6pt}{}
\be
\int^{+\infty}_{-\infty} d\rho f^2(\rho) \left\langle \left(\Tren_{\tau \tau} \left(\frac{\partial \tau}{\partial \lambda}\right)^2+\Tren_{\rho \rho} \right) \right\rangle_{\psi} \geq - \frac{4B}{G_N} \int^{+\infty}_{-\infty} d \rho \left(f'(\rho)\right)^2 \,.
\ee
Using Equation~\eqref{eqn:stresstensrho}, we have
\vspace{-6pt}{}
\be
\int^{\infty}_{0} d\rho \, f(\rho)^2 \frac{2\ell^2}{(1+\rho^2)^2} \langle T_{tt}\rangle  \geq  -\frac{4B}{G_N} \int^{\infty}_{0} d\rho f'(\rho)^2 \,.
\ee
We note that this derivation differs slightly from Ref.~\cite{Freivogel:2020hiz}. In~particular, Equation~(57) from that reference holds only on the geodesic and not in general. Inside the wormhole throat, the~stress--energy tensor is a constant given by Equation~\eqref{eqn:wormstress}, and using Equation~\eqref{eqn:rexpr}, we can define
\be
\label{eqn:cc}
C \equiv \frac{|\langle T_{--} \rangle | G_N}{4B}=\frac{g^2}{24 B \pi^3 q } A \,,
\ee
using $\ell_{\text{pl}}=\sqrt{G_N}$. Using Equation~\eqref{eqn:cc}, we have
\vspace{-6pt}{}
\begin{equation}
	\label{eqn:snecwithc}
	C \int^{\infty}_{0} d\rho f(\rho)^2 \frac{1}{(1+\rho^2)^2}  \leq   \int^{\infty}_{0} d\rho f'(\rho)^2 \,.
\end{equation}
The function $f(\rho)$ is undefined up to this point. It needs to be a function of compact support or a function that decreases fast at $\pm \infty$ in order to have a finite lower bound. One could try to solve the variational problem to find the optimal function. However, it is quite complicated, and as $B$ and $q$ are not specified, it is more instructive to examine their role with a simple choice of function. For
\vspace{-6pt}{}
\be
f(\rho)^2=\frac{1}{\sigma} \sqrt{\frac{2}{\pi}} e^{-\frac{\rho^2}{2\sigma^2}} \,,
\ee
with a normalized Gaussian function of width $\sigma$, Equation~\eqref{eqn:snecwithc} becomes
\vspace{-6pt}{}
\begin{equation}
	\label{MaldBoundGaussian2}
	C\int^{\infty}_{0} d\rho ~ e^{-\frac{\rho^2}{2\sigma^2}}  ~ \frac{1}{(1+\rho^2)^2} \leq \frac{1}{\sigma^4}   \int^{\infty}_{0} d\rho ~ \frac{\rho^2}{4}~  e^{-\frac{\rho^2}{2\sigma^2}}   \,.
\end{equation}
As in Ref.~\cite{Freivogel:2020hiz}, a~computation of the integrals gives
\vspace{-6pt}{}
\be
\label{eqn:expgauss}
C \leq  \left(2+\sigma^{-1} e^{\frac{1}{2\sigma^2}} \sqrt{2\pi} (\sigma^2-1) \erf{\left[\frac{1}{\sqrt{2} \sigma}\right]} \right)^{-1} \,,
\ee
where $\erf$ is the error function \endnote{The expression differs slightly from Ref.~\cite{Freivogel:2020hiz} due to the different definition of the Gaussian.}.

Setting $\sigma=\rho_0$ as the~length of the achronal segment, 
 we have
\vspace{-6pt}{}
\be
Bq \gtrapprox 1.3 \times 10^{-2} \,.
\ee
Here, we also set $Ag^2$ to $1$. If~$B$ has its maximum value of $1/32\pi^2$, we need $q \lessapprox 1$ to saturate the SNEC. But,~$q \gg 1$ so that the wormhole is stabilized. Then, for this maximum value, the SNEC is not saturated. We need $B \lessapprox  1/q$ to have that for the $\rho_0=4.13$ value. 

\textls[-25]{Keeping the maximum $B$ value and setting $q\sim 10^2$, we can find how long the achronal segment needs to be so that the SNEC is saturated. Numerically, we find the value $\rho_0= \sigma \sim 293$.}

\subsection{DSNEC}

Next, we examine the DSNEC bound in the case of the Maldacena--Milekhin--Popov wormhole. We will use the bound \eqref{eq:DSNECdimless}, which is---in terms of dimensionless \mbox{parameters---$s^\pm$:}
\vspace{-6pt}{}
\bea
&& \int  \frac{d^2x ^{\pm}}{\delta^+ \delta^-} \mc F(s^+,s^-)^2\langle \Tren_{--}\rangle_{\psi}\geq \nonumber \\
&& \qquad -\frac{16}{81\pi^2}\frac{1}{\delta^+ (\delta^-)^3} \left(\int ds^+ (\mc F_+''(s^+))^2\right)^{1/4} \left(\int ds^- (\mc F_-''(s^-))^2\right)^{3/4} \,. 
\eea
Now, we need to recall the discussion in Section~\ref{sub:nullqei} regarding curvature. Neither the SNEC nor the DNSEC have been proven for curved spacetimes. In~the SNEC section, we assumed that a SNEC-type bound holds for spacetimes with curvature, perhaps with a different $B$. However, as~was discussed, in~the DSNEC case, the~bound for Minkowski spacetime is not generally covariant. Additionally, it is unclear what integrating over two null directions means unless they are both less that the radius of curvature. Thus, for the DNEC, we will use a flat spacetime metric to describe short distances in the throat of the wormhole:
\vspace{-6pt}{}
\be
ds^2=-r_e^2 ds^- ds^+ \,,
\ee
where $s^-=\tau-\rho$ and $s^+=\tau+\rho$ for $\rho+1 \approx 1$.

We will use similar considerations as used for the SNEC. We define
\vspace{-3pt}{}
\be
\tilde{C}\equiv \frac{|\langle T_{--} \rangle | 81 \pi^2}{16}=\frac{|\langle T_{tt} \rangle | 81 \pi^2}{8}=\frac{27}{32} \frac{q A}{ r_e^2 \ell^2}   \,.
\ee
As in the previous section, we pick Gaussian functions for $\mc F_-$ and $\mc F_+$. Then, we have
\vspace{-6pt}{}
\be
\tilde{C} \leq \frac{3}{(\delta^-)^3 \delta^+} \,.
\ee
We have $\delta^-=\ell$, which is the~length of the wormhole for $\rho=\rho_0$. The~length for the other direction is taken as the $AdS_2$ radius, $r_e$.

Taking $A$ of order $1$, to~violate the DSNEC, we need
\vspace{-6pt}{}
\be
q \gtrapprox \frac{r_e}{\ell} \,.
\ee
As $r_e \ll \ell$, the DSNEC is easily~violated.

One might wonder why there is a difference between the SNEC and the DNEC. To~examine that, we look at the schematic forms of the bounds
\vspace{-6pt}{}
\be
\langle T_{--} \rangle_{\text{SNEC}} \geq -\frac{4B}{\ell_\text{pl}^2 \ell^2} \,, \qquad \langle T_{--} \rangle_{\text{DSNEC}} \geq -\frac{N}{\ell^3 r_e} \,.
\ee
Taking $B\sim N \sim 1$, it is immediately obvious that the DSNEC bound is much stricter in this case. Of~course, we should take these results with a grain of salt: the SNEC and DSNEC bounds are for free massless scalars on Minkowski. They are expected to be modified when they incorporate curvature and, more importantly, interactions to accurately be applied to this wormhole~case.

\subsection{Two-Dimensional~Bound}

In Ref.~\cite{Freivogel:2020hiz}, it was shown that the~two-dimensional bound of Fewster and Hollands~\cite{Fewster:2004nj} given in Equation~\eqref{eqn:conf2d} is saturated for the wormhole~\cite{Maldacena:2018gjk}. This bound holds for spacetimes globally conformal to Minkowski, as shown in Ref.~\cite{Freivogel:2020hiz} following the work of Flanagan~\cite{Flanagan:2002bd}.

While $AdS_2$ is not globally conformal to Minkowski, the bound can be used far from the boundaries in the wormhole case. We have
\vspace{-6pt}{}
\begin{equation}
	\label{eqn:2dsnec}
	\int^{\infty}_{0} d\rho f(\rho)^2 \frac{1}{(1+\rho^2)^2} |\langle T^{(2)}_{--} \rangle | \leq  \frac{c}{12\pi} \int^{\infty}_{0} d\rho f'(\rho)^2 \,,
\end{equation}
where $T^{(2)}$ is the two-dimensional stress--energy tensor given by
\vspace{-6pt}{}
\begin{equation}
	T^{(2)}_{--} = r^2_e ~  T_{--} \propto q \,.
\end{equation}
Then, Equation~\eqref{eqn:2dsnec} becomes
\begin{equation}
	\label{q_over_c}
	\frac{q}{c} \int^{\infty}_{0} d\rho f(\rho)^2 \frac{1}{(1+\rho^2)^2} \leq   \int^{\infty}_{0} d\rho f'(\rho)^2 \,.
\end{equation}
In the CFTs, the number of fields should also be given by $c$, so $q/c$ is of order one. Similarly to the previous subsections, after~the integrals are computed, we have
\vspace{-6pt}{}
\be
\label{eqn:expgauss 2d}
\frac{q}{c} \leq \sqrt{2} \left(2+\sigma^{-1} e^{2/\sigma^2} \sqrt{\pi} (\sigma^2-4) \erf{[\sqrt{2}/\sigma]} \right)^{-1} \,.
\ee
It was noticed that in both cases where $\sigma \ll 1$ that
\vspace{-6pt}{}
\be
\frac{q}{c} \leq \frac{1}{\sqrt{2} \sigma^2} +\mathcal{O}(\sigma^4) \,,
\ee
and for $\sigma \gg 1$ 
\be
\frac{q}{c} \leq \frac{\sqrt{2}}{\sqrt{\pi} \sigma}+\mathcal{O}\left(\frac{1}{\sigma^3}\right) \,,
\ee
so the bound can be saturated. This can be seen, as $q/c$ is an order one number, and~$\sigma < 4.13$.

\section{Discussion}
\label{sec:discussion}

In this review, I examined various constraints imposed to wormholes by derived energy conditions. In~the case of `short' wormholes that allow causality violations, the~strongest restriction is the achronal ANEC. No counterexamples to the self-consistent achronal ANEC have been found in the context of semiclassical gravity, and it has been shown~\cite{Graham:2007va} that it is sufficient to prevent causality~violations. 

What possibilities remain for wormholes? Assuming that the spacetime is generic, only three exist: one is that the spacetime is not asymptotically flat. This is possible with some time machine constructions; however, those often assume infinite energy such as the Mallette time machine~\cite{mallett2003gravitational}. Perhaps the constraints from QEIs could provide further restrictions on those types of constructions. However, most wormhole constructions assume asymptotic flatness in order for observers to enter the mouths. A~different situation is that of asymptotically AdS spacetimes, which is of special interest due to the AdS/CFT correspondence. It seems possible that some of the theorems for asymptotically flat spacetimes could carry over to asymptotically AdS spacetimes, but that remains to be~seen.

The second is that we are not in the realm of semiclassical gravity. A~quantum gravity regime wormhole with a throat of Planck scale size~\cite{Hochberg:1996ee} would not be traversable, so this is not usually considered an option. Alternative theories of gravity such as $f(R)$ are often the framework of wormhole solutions. Alternately, hypothetical matter fields that could violate the achronal ANEC can also be considered such as `phantom energy' (see~\cite{Lobo:2016zle} for references within). But,~those have not been~detected. 

The third possibility is that the geodesics passing through the wormhole are not achronal. As~was discussed, the ANEC can be violated in this case. However, those `long' wormholes do not allow for causality violations, and they are not much use for interstellar travel. In~the case of those wormholes, such as the one described in Ref.~\cite{Maldacena:2018gjk}, quantum energy inequalities (QEIs) can provide restrictions. This paper discussed restrictions from null QEIs and presented a new result in the case of the double smeared null energy condition (DSNEC). It seems that the DNSEC could provide further restrictions to the length of long wormholes, but more work is required to verify~that. 

A significant obstacle is that the QEIs used are all for Minkowski spacetime. In~the case of timelike ones, a~bound that incorporates curvature exists~\cite{Kontou:2014tha}, so an application is possible, albiet technically challenging. In~the case of the DSNEC, while there is a way to incorporate curvature, such an explicit bound has not been derived yet. It would be of interest to examine wormhole solutions with such bounds on curved spacetimes. Finally, QEIs for self-interacting fields do not exist, except for in two dimensions (see Section~\ref{sub:QEIs} for references). 
 Such bounds might differ significantly from the ones used, and as wormhole solutions often utilize self-interacting quantum fields, this means that the results are~unknown.

\vspace{6pt} 




\funding{This research received no external~funding.}

\dataavailability{Data are contained within the article. }

\acknowledgments{The author would like to thank Chris Fewster, Ben Freivogel, and Dimitrios Krommydas for useful discussions. The~idea that the DSNEC might be violated in the case of the Maldacena--Milekhin--Popov wormhole was first presented by Ben Freivogel during the conference ``Energy conditions in quantum field theory'' in Leipzig, Germany, 2022.}

\conflictsofinterest{The author declares no conflicts of~interest.}

\begin{adjustwidth}{-\extralength}{0cm}
\printendnotes[custom] 

\reftitle{References}



\PublishersNote{}
\end{adjustwidth}
\end{document}